%% file: main_manuscript-2.tex
\documentclass{nature}
\usepackage{amsmath}
\usepackage{textcomp}
\usepackage{lscape}  
\usepackage{graphicx}
\usepackage{color}
\usepackage{cite}
\usepackage{float}
\usepackage{ccaption}
\makeatletter
\let\saved@includegraphics\includegraphics
\AtBeginDocument{\let\includegraphics\saved@includegraphics}

\makeatother

  \title{Multi-bit Ferroelectric-NAND for High-throughput Massive Database Search}

 \author{Prasanna Venkatesan\textsuperscript{1,*}, Tanvir H. Pantha\textsuperscript{2}, Po-Kai Hsu\textsuperscript{1}, Sumukh Pinge\textsuperscript{3}, Zheyu Li\textsuperscript{3}, Chinsung Park\textsuperscript{1}, Priyankka Ravikumar\textsuperscript{1}, Hari Jayasankar\textsuperscript{1}, Lance Fernandes\textsuperscript{1}, Weihong Xu\textsuperscript{3}, Zihan Xia\textsuperscript{3}, Flavio Ponzina\textsuperscript{3}, Keming Fan\textsuperscript{3}, Amrit Garlapati\textsuperscript{1}, Huy Tran\textsuperscript{1}, Taeyoung Song\textsuperscript{1}, Mengkun Tian\textsuperscript{4}, Hang Chen\textsuperscript{5}, Winston Chern\textsuperscript{1}, Kijoon Kim\textsuperscript{6}, Kwangyou Seo\textsuperscript{6}, Suhwan Lim\textsuperscript{6}, Kwangsoo Kim\textsuperscript{6}, Wanki Kim\textsuperscript{6}, Daewon Ha\textsuperscript{6}, Duygu Kuzum\textsuperscript{3}, Shimeng Yu\textsuperscript{1}, Mingu Kang\textsuperscript{3}, Tajana Rosing\textsuperscript{3}, Sourav Dutta\textsuperscript{2}, Asif Khan\textsuperscript{1,7,*}}

\begin{document}

 \maketitle
\vspace{-.3in}
\begin{affiliations}
\item School of Electrical and Computer Engineering, Georgia Institute of Technology, Atlanta, GA, USA.

\item School of Electrical and Computer Engineering, University of Texas Dallas, 
Richardson, TX, USA.

\item School of Electrical and Computer Engineering, University of California San Diego, San Diego, CA, USA.

\item Institute of Materials, Georgia Institute of Technology, Atlanta, GA, USA.

\item Institute of Matter and Systems, Georgia Institute of Technology, Atlanta, GA, USA.

\item Advanced Device Research Laboratory, Semiconductor Research Center, Samsung Electronics Co., Ltd., South Korea.

\item School of Materials Science and Engineering, Georgia Institute of Technology, Atlanta, GA, USA.

\item[\textsuperscript{*}] Corresponding authors. Email: pravindran6@gatech.edu, akhan40@gatech.edu.  

\end{affiliations}

\newpage
\begin{abstract}


The growing demand for large-scale database search in data-intensive applications—ranging from proteomics to autonomous systems—has exposed fundamental limitations in von Neumann architectures due to memory bandwidth and energy constraints. Hyperdimensional (HD) computing offers a robust and parallelizable framework for such tasks, but its practical implementation remains challenged by high memory demands. Ultra-high-density, energy-efficient ferroelectric NAND (FE-NAND) memory provides a potential solution by enabling \textit{in-situ} computation. We fabricate quad-level FE-NAND strings with wide memory windows, disturb resilience, and robust retention. Using these \textcolor{black}{planar FE-NAND strings as building blocks,} \textcolor{black}{we experimentally demonstrate \textit{in-situ} multi-level cell (MLC) dot product operations at the single-cell level and, using experimentally calibrated physics-based simulations, demonstrate} Hamming similarity calculations between reference and query hypervectors (HV). This platform leverages the inherent error tolerance of HD computing to achieve $>$90\% search accuracy even at high logic levels (TLC, QLC). When benchmarked on Open Modification Search (OMS) tasks in proteomics with TB-scale datasets, our system shows nearly 1,000$\times$ speedup and over 10,000$\times$ energy efficiency improvement compared to incumbent solutions. These results establish FE-NAND as a viable in-storage compute architecture for large-scale, high-dimensional data processing.
\newpage


\end{abstract}

\newpage

\section{Introduction}


The exponential growth of data across domains, from large language models and real-time autonomous systems to proteomics and genomics, has made large-scale database search a fundamental yet increasingly challenging task. Traditional computing architectures, based on von Neumann principles, suffer from limited bandwidth between memory and compute units, leading to performance bottlenecks and high energy costs when processing massive datasets. These challenges are especially pronounced in similarity-based search tasks, which often involve high-dimensional data representations and require repeated, memory-intensive computations. Hyperdimensional (HD) computing offers a promising alternative by enabling fast, parallel, and inherently error-tolerant search through distributed representations\cite{imani2019searchd,bittremieux2019extremely}.  HD computing is a cutting-edge computational paradigm that utilizes high-dimensional vectors, or hypervectors, to represent and process information. Inspired by the human brain's data handling for pattern-oriented computations, HD computing offers robust, noise-tolerant, energy-efficient and parallelized solutions for complex tasks involving pattern matching in large datasets. Its robustness and capability for approximate matching make it particularly well-suited for large-scale, noisy, or real-time data environments.


Data-driven discovery is key to many emerging applications. The volume of data is far outstripping the ability to store and process it. For example, the data movement and computational bottlenecks related to processing petabytes of data limit the first stage of drug discovery to over 6.5 years while still facing downstream failure rates of over 90\%\cite{dowden2019trends}. Better understanding biological processes could enable us to best select compounds to put through clinical trials at upto 100x shorter times. These biological processes are analyzed using ``omics" disciplines - genomics (DNA) and transcriptomics (RNA), which use sequencers, along with proteomics (proteins) and metabolomics (metabolites), which rely on Mass Spectrometry. Virtual drug screening is extensively used to systematically search compound libraries for small molecules that bind to a target, and thus can be used to develop the next generation personalized medicine.  The key challenge with such workloads is the sheer size and rate of data generation. For example, UCSD’s protein mass spectrometry database is at 1055 TB and growing\cite{MassIVE2024}. Modern mass spectrometers generate millions of spectra during each 30 minute experiment, creating hundreds of gigabytes of spectral data to process. Searching for protein matches in a large scale mass spectrometry database \textit{via} Open Modification Search is one of the key steps in the drug discovery pipeline\cite{bittremieux2019extremely} (Fig. 1(e)).

HD computing algorithms for OMS applications have been developed over the last decade and provide robust and comprehensive analysis but their efficiency on CPUs is low due to fundamental efficiency barriers in von-Neumann architectures, where excessive data movement consumes more than 60\% of runtime and limited parallelism bottlenecks bulk bitwise operations. GPU tools, like HyperOMS\cite{kang2022massively}, improve OMS efficiency, however, they face scalability challenges due to the large memory footprint incurred by memory-intensive HD computing primitives. The two computing primitives – encoding and search are memory bandwidth limited, with search accounting for nearly 85\% of memory utilization. While near-memory and in-memory computing approaches leveraging phase-change memory arrays and resistive crossbar arrays enable low-energy, high-speed processing, they face limitations in scalability and memory density. In contrast, in-storage computation using 3D NAND, where data is processed in NAND arrays without being moved out, offers a more scalable path with significantly higher capacity, making it well-suited for processing large datasets.

\textcolor{black}{Prior works have explored FeFET-based edge detection using binary operations~\cite{reis2023low}, ferroelectric NAND for bayesian neural networks\cite{song2025ferroelectric}, 3D CTF NAND for hyperdimensional genome sequencing using near-memory circuits~\cite{wu2024heterogeneous}, and sensor-coupled NAND HDC architectures for image processing~\cite{salamat2021senhdc}. In contrast, the present work demonstrates \textit{in-situ} multi-level cell (MLC) dot product computation directly within a ferroelectric NAND string with a QLC-compatible memory window, targeting TB-scale proteomics database search---a workload where in-storage compute density is the dominant performance driver.}


Charge-trap flash (CTF) NAND has served as the backbone of large-scale data storage over the past decade, pioneering 3D memory architectures through the integration of vertical channel transistors. By leveraging stacked wordline layers and macaroni-shaped charge-trapping cells, CTF NAND has enabled significant advances in bit-areal density scaling\cite{tanaka2007bit,fukuzumi2007optimal,park2014three}. Continued scaling has been achieved through logic scaling (storing multiple bits per cell), lateral scaling (\textit{via} optimized memory hole layouts), and vertical scaling (through reduced z-pitch and increased layer counts), resulting in more than a tenfold increase in storage density over the last decade\cite{goda2021recent,goda2024nand}. However, further vertical stacking is increasingly constrained by reliability and packaging limitations\cite{han2023fundamental}. Shrinking the z-pitch introduces retention challenges due to lateral charge migration, while increasing layer count at fixed z-pitch leads to total stack heights that exceed package tolerances. Recent studies have proposed integrating ferroelectric materials into the NAND gate stack to address these limitations, offering a pathway for continued vertical scaling by enabling tighter z-pitch without compromising reliability\cite{das2023experimental,venkatesan2024disturb,yoon2023qlc,das2024ferroelectric, yoo2025ferroelectric}. As shown in Fig. 1(a), the integration of ferroelectrics into NAND architectures (FE-NAND) provides a pathway towards achieving over 1000-layer stacks and storage densities exceeding 100~Gb/mm$^2$.

By unifying the ultra-high-density potential of ferroelectric materials with the in-storage compute capabilities of 3D NAND architectures, FE-NAND provides a scalable and efficient platform for data-centric computation. \textcolor{black}{While MFIFS gate stacks have been previously reported~\cite{das2023experimental,das2024ferroelectric}, the present work leverages such structures for \textit{in-situ} multi-level compute operations within FE-NAND strings, demonstrating MLC dot product computation and Hamming similarity search for in-storage hyperdimensional computing.} In this work, we investigate FE-NAND as a compute-capable storage solution for HD computing, using protein identification \textit{via} Open Modification Search (OMS) as a representative application. \textcolor{black}{We design and fabricate planar FE-NAND strings on SOI as building blocks. The MFIFS gate-stack physics---memory window, retention, disturb resilience, and endurance---demonstrated on these planar devices are directly transferable to 3D vertical NAND architectures, where the same conformal ALD gate stack wraps around the cylindrical channel. System-level performance is projected using experimentally calibrated device models within a 3D NAND architectural framework.} We demonstrate key memory properties including disturb-free read, high retention, and endurance. Leveraging the multi-level programmability of FE-NAND, we \textcolor{black}{experimentally demonstrate \textit{in-situ} bitwise MLC dot product at the single-cell level and, through experimentally calibrated simulations,} Hamming distance calculations between hypervectors, both fundamental operations in database search using HD computing. \textcolor{black}{Here, the bitwise dot product is verified experimentally at the single-cell level within fabricated FE-NAND strings, while the array-level Hamming similarity and system-level benchmarks are obtained from experimentally calibrated, variation- and degradation-aware physics-based simulations.} This enables scalable, high-throughput similarity search without data movement between memory and processor. By aligning the intrinsic parallelism and error tolerance of HD computing with the density and efficiency of FE-NAND, our platform achieves accurate, energy-efficient in-storage search across large-scale datasets.

\section{Ferroelectric NAND string and device characteristics}

\textcolor{black}{Ferroelectric NAND (FE-NAND) strings were fabricated on a silicon-on-insulator (SOI) platform as planar devices, as detailed in the Methods section and Supplementary Note~1. The planar geometry provides direct access to the gate-stack physics that governs device behavior in both planar and vertical configurations.} Fig.\ref{device}(b) shows a false-colored scanning electron micrograph (SEM) of the FE-NAND strings with four wordlines (highlighted in blue). In a conventional approach, a single Hf\textsubscript{0.5}Zr\textsubscript{0.5}O\textsubscript{2} (HZO) ferroelectric layer is used as the gate dielectric. This results in a limited memory window (MW), typically around 3~V, mainly constrained by the coercive voltage (2$V_C$) of the single Hf\textsubscript{0.5}Zr\textsubscript{0.5}O\textsubscript{2} (HZO) ferroelectric layer. To overcome this limitation, we introduce a dielectric into the ferroelectric gate stack. We insert a thin 3~nm Al\textsubscript{2}O\textsubscript{3} dielectric layer sandwiched between two ferroelectric HZO layers. This allows a threefold enhancement of the MW without increasing the overall stack thickness. The resulting $I_D$–$V_G$ characteristics, shown in Fig.\ref{device}(d), confirms this enhancement. Cross-sectional transmission electron microscopy (TEM) images and energy-dispersive X-ray spectroscopy (EDS) elemental maps of the modified gate stack are presented in Fig.\ref{device}(c), clearly illustrating the inserted dielectric layer. As shown in Fig.\ref{device}(e), the MW–$V_{write}$ behavior of both devices exhibits similar trends at low voltages; however, the standard FEFET saturates at an MW of 2.4~V beyond 11~V, while the engineered stack shows a sharp increase, reaching a MW of 7.5~V at 14~V, enabling quad-level cell (QLC) operation. This enhanced MW can be attributed  to charge trapping phenomena at the ferroelectric–dielectric interface, as reported in another study\cite{das2023experimental}. A brief discussion on multi-level logic compatibility is included in Supplementary Note 9.

FE-NAND strings operate by applying a pass voltage higher than the threshold voltage ($V_T$) of the erase (ERS) state at all unselected wordlines, thereby turning on the channel for bitline access. However, repeated application of such positive pass pulses can induce $V_T$ shifts over time due to electron trapping, analogous to positive bias temperature instability (PBTI). To mitigate this effect, we implement a periodic refresh scheme that combines a negative pulse to detrap accumulated electrons, followed by a program (PGM) pulse to restore the intended memory state as shown in Supplementary Fig. S2. This scheme effectively stabilizes the threshold voltages of all four states, with $<$ 4\% shift across disturb cycles relative to the total memory window\cite{venkatesan2024disturb}.

While introducing a dielectric layer near the gate interface can enhance the MW, it also introduces retention concerns associated with detrapping of the MW enhancing charges at the ferroelectric-dielectric interface—causing the initial MW to degrade with increasing temperature and results in significant retention loss even at room temperature\cite{venkatesan2024demonstration}. In contrast, our approach of embedding the dielectric layer within the middle of the ferroelectric stack enables robust memory performance (Supplementary Fig. S2). Specifically, we observe $<$1\% retention loss after $10^4$ seconds. Additionally, we demonstrate an endurance of $10^4$ program/erase cycles, with a stable MW maintained through at least 3000 cycles as shown in Supplementary Fig. S2. \textcolor{black}{The $10^4$ P/E endurance of the MFIFS gate stack, limited by charge trapping at the ferroelectric--dielectric interface, is comparable to the $10^3$--$10^4$ P/E specification of commercial QLC/TLC NAND and is sufficient for the read-dominant in-storage compute workloads demonstrated here. While improved endurance through interface engineering has been demonstrated in MFIS FeFETs, improving MFIFS endurance is an important direction for future work \cite{venkatesan2025pushing,ravikumar2025experimental}.}

\textcolor{black}{Importantly, the FE-NAND in-storage compute scheme uses conventional digital NAND programming (ISPP with verify) and multi-step digital readout, rather than analog weight programming. The programming precision requirements are therefore identical to those of commercial QLC/TLC/MLC NAND---unlike analog in-memory compute approaches where precise conductance programming is a key challenge~\cite{lee2024analog}. Selective cell programming in a crossed BL/WL array follows the standard NAND inhibit scheme. Unselected BLs are pre-charged to boost the channel potential below the ferroelectric coercive threshold, suppressing program disturb.  A disturb mitigation scheme---demonstrating $<$4\% $V_T$ shift under repeated $V_{pass}$ stress with periodic refresh---has been validated in our prior work \cite{venkatesan2024disturb}, achieving disturb levels comparable to CTF NAND ($<$3\%).}

Historically, FE-NAND devices have been constrained by limited memory window and lack of reliable multi-level operation. In this work, we overcome these limitations, demonstrating disturb-resilient behavior comparable to CTF NAND, record-high retention of over 10 years, lower write voltages, and a QLC-compatible memory window. These advances establish FE-NAND not only as a scalable storage solution, but also as a robust platform for executing in-storage computations. Building on this foundation, we next demonstrate the execution of core HD computing operations, specifically, \textit{in-situ} multi-level dot product and Hamming distance calculations, directly within the FE-NAND array.

\section{\textit{In-situ} multi-bit dot product in FE-NAND for hyperdimensional search}

In HD computing, hyperdimensional algebra is leveraged to encode the dataset by distributing the information across all dimensions, while preserving semantic relationships. Search operations rely on distance metrics, such as Hamming similarity, to compare query hypervectors with reference hypervectors. In our implementation, reference hypervectors ($\hat{W}$) are stored in the FE-NAND array, with one hypervector mapped per row/wordline.  The query hypervector ($\hat{A}$) is applied along the bitlines as a voltage level and a \textit{bitwise} dot product between the query and reference vectors is performed to compute similarity. This dot product step is a critical component in Open Modification Search (OMS) workflows and underpins the associative retrieval operations in HDC. \textcolor{black}{The bitwise MLC dot product is demonstrated experimentally at the single-cell level within fabricated FE-NAND strings on SOI. The query bit is passed as a BL voltage and the reference bit is stored as a multi-level $V_T$ state. The 4-step read process (using $V_A$, $V_B$, $V_C$) digitizes the two-bit product at each BL position. This per-cell operation is the fundamental compute primitive for full dot product and Hamming similarity computation in FE-NAND arrays.}

To compute the \textit{bitwise} dot product $B = \hat{A} \cdot \hat{W}$, the two bits of the query vector ($\hat{A}_{2j}$, $\hat{A}_{2j+1}$) are sequentially passed at each bitline (BL), corresponding to two bits of the reference hypervector ($\hat{W}_{2j}$, $\hat{W}_{2j+1}$) stored in the cell as a multi-level threshold voltage as shown in Fig.~\ref{array}(b). As described in Fig. \ref{array}, the resulting outputs are computed as $B_{2j} = \hat{A}{2j} \cdot \hat{W}_{2j}$ and $B_{2j+1} = \hat{A}_{2j+1} \cdot \hat{W}_{2j+1}$, forming the two-bit output $B_{2j}B_{2j+1}$ at the $j^{th}$ BL. 
The output at each BL, $B_{2j}B_{2j+1}$, is read through a four-step read process using different read reference voltages ($V_A$, $V_B$, $V_C$) applied at the gate. 
As shown in Fig.~\ref{array}(b), $V_B$ distinguishes between the left and right halves of the threshold voltage range to determine the LSB, while $V_A$ and $V_C$ subsequently divide the range into quarters to determine the MSB. Therefore, the LSB of the dot product,  $B_{2j+1}$, is obtained from $I_{b1}$, the BL current with $V_B$ as the read voltage and $A_{2j+1}$ as the BL voltage. $A_{2j}$ is passed at the BL for the next three reads with read voltages $V_B$, $V_A$ and $V_C$ to obtain the three BL currents $I_{b2}$, $I_a$ and $I_c$, respectively. 
Therefore, the MSB of the dot product, $B_{2j}$, is computed from the digitized currents as $I_{c}\cdot(I_{a}\odot I_{b2})$, where $I_{a}\odot I_{b2} = I_{a}\cdot I_{b2} + \overline{I_{a}}\cdot \overline{I_{b2}}$. We illustrate the MLC dot product calculation for two representative cases, $W_{ij} = 11$ and $A_j = 01$ and \textcolor{black}{$W_{ij} = 01$} and $A_j = 10$ in Fig. \ref{array}(c,d) showing the current transients and the computed output. 

By repeating the dot product operation across all combinations of $A_j$ ($00$, $01$, $10$, $11$) and $W_{ij}$ ($00$, $01$, $10$, $11$), the truth table for \textit{bitwise} MLC dot product was experimentally reproduced as shown in Fig. \ref{array}(e). The bitline outputs from each read operation were digitized using a 1~nA bitline current threshold during post-processing.

\begin{figure*}
\centering
\includegraphics[width = 0.9\textwidth]{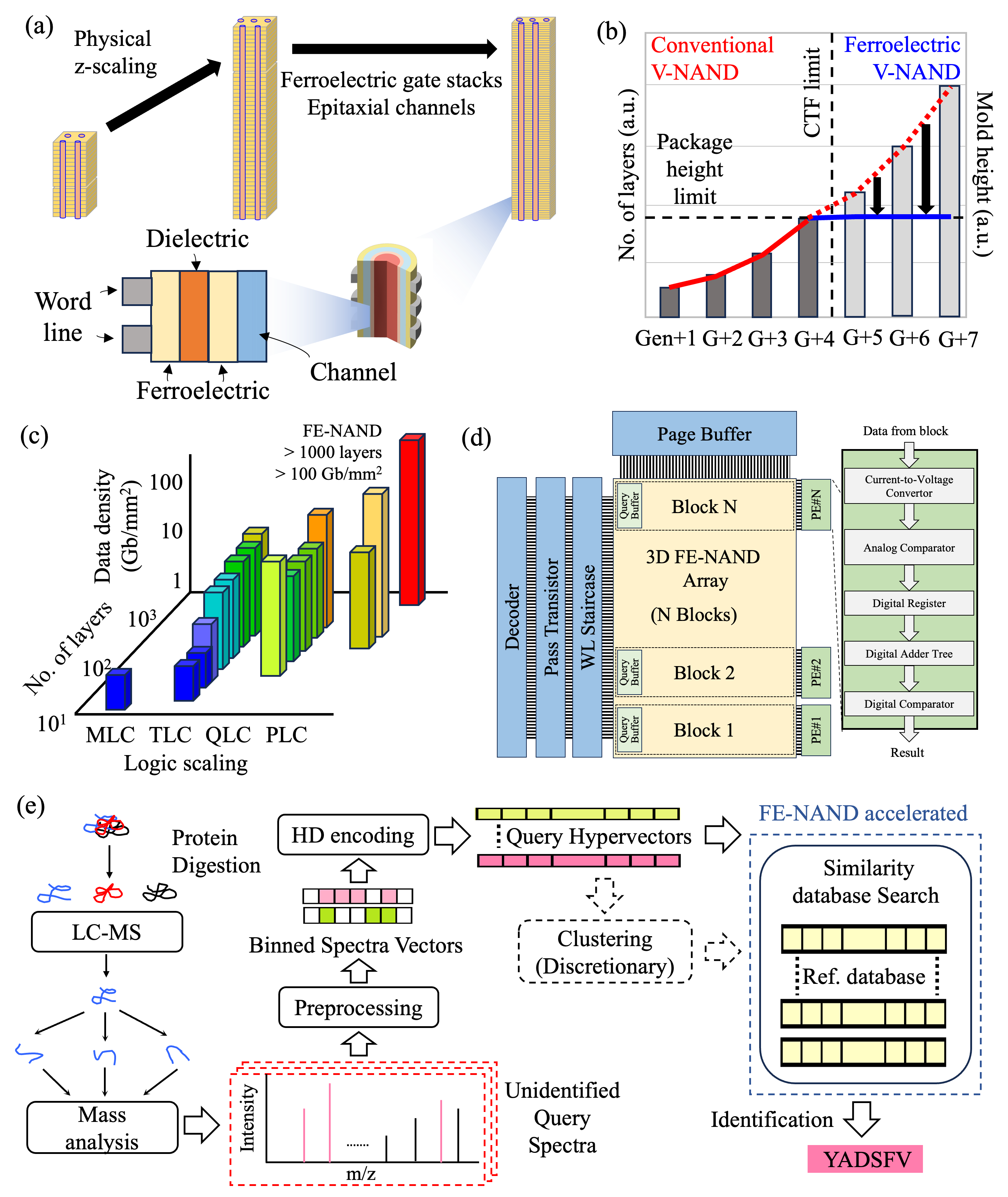}
\vspace{-0.2in}
\caption{ (a) Increasing the number of wordlines in 3D has been enabled by physical z-scaling. Saturation in z-pitch scaling coupled with package height limitations have limited further z-height increase. Innovations in 3D NAND design and processing, including Ferroelectric gate stacks, epitaxial channels and cryogenic etching, provide a pathway towards increase WL density.  (b) As the number of layers increase in NAND solutions, the total z-height of the NAND solutions is approaching the package height\cite{han2023fundamental}. Importantly, further z-pitch scaling is limited by reliability challenges like retention loss and high operating voltages. (c) A summary of data density scaling in 3D NAND over the last decade showing increasing data density through increased WL count and logic scaling\cite{jung202413,park2014three,kim20221tb,khakifirooz20231,kim202328,hu2022512gb}. Ferroelectric NAND has emerged as a solution for beyond 1000 layer NAND by enabling further z-pitch scaling due to its resilience against the reliability challenges faced by conventional NAND\cite{das2023experimental,yoon2023qlc,venkatesan2024disturb}. (d) Framework for in-storage compute in 3D FE-NAND. (e) Workflow for FE-NAND accelerated-open modification search (OMS) for protein identification\cite{bittremieux2019extremely}.  } \label{intro}
\end{figure*}

\begin{figure*}[!b]
\centering
\includegraphics[width = \textwidth]{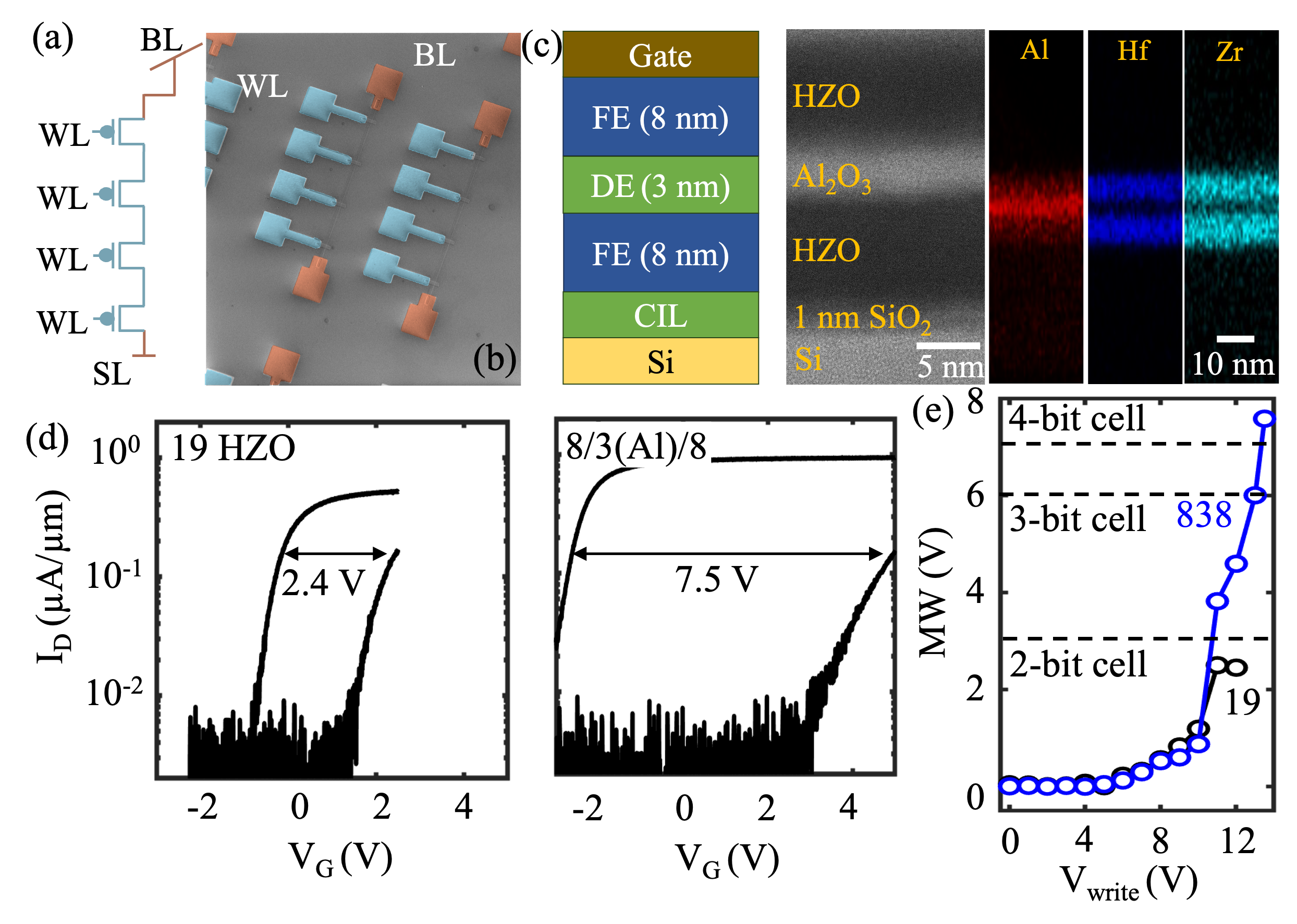}
\vspace{-0.2in}
\caption{(a) Schematic of a NAND string with four wordlines. (b) A SEM of the fabricated ferroelectric NAND strings. (c) TEM of the band-engineered 8/3/8 gate stack showing the 3 nm Al\textsubscript{2}O\textsubscript{3} insert along with EDS maps showing the elemental composition. \textcolor{black}{CIL: channel interface layer (SiO$_x$ between the Si channel and ferroelectric gate stack).} (d) Transfer characteristics of a single cell with reference 19 nm HZO gate stack and 8/3(Al)/8 gate stack showing a QLC compatible MW of 7.5 V, three times that of the reference FEFET. (e) Evolution of the MW with write voltage showing the steep increase in MW in 8/3/8 FEFETs. \textcolor{black}{The four threshold-voltage levels demonstrated here correspond to four distinct states, i.e. 2-bit (MLC) operation (four states).}}
\label{device}
\end{figure*}

\begin{figure*}
\centering
\includegraphics[width = 0.85\textwidth]{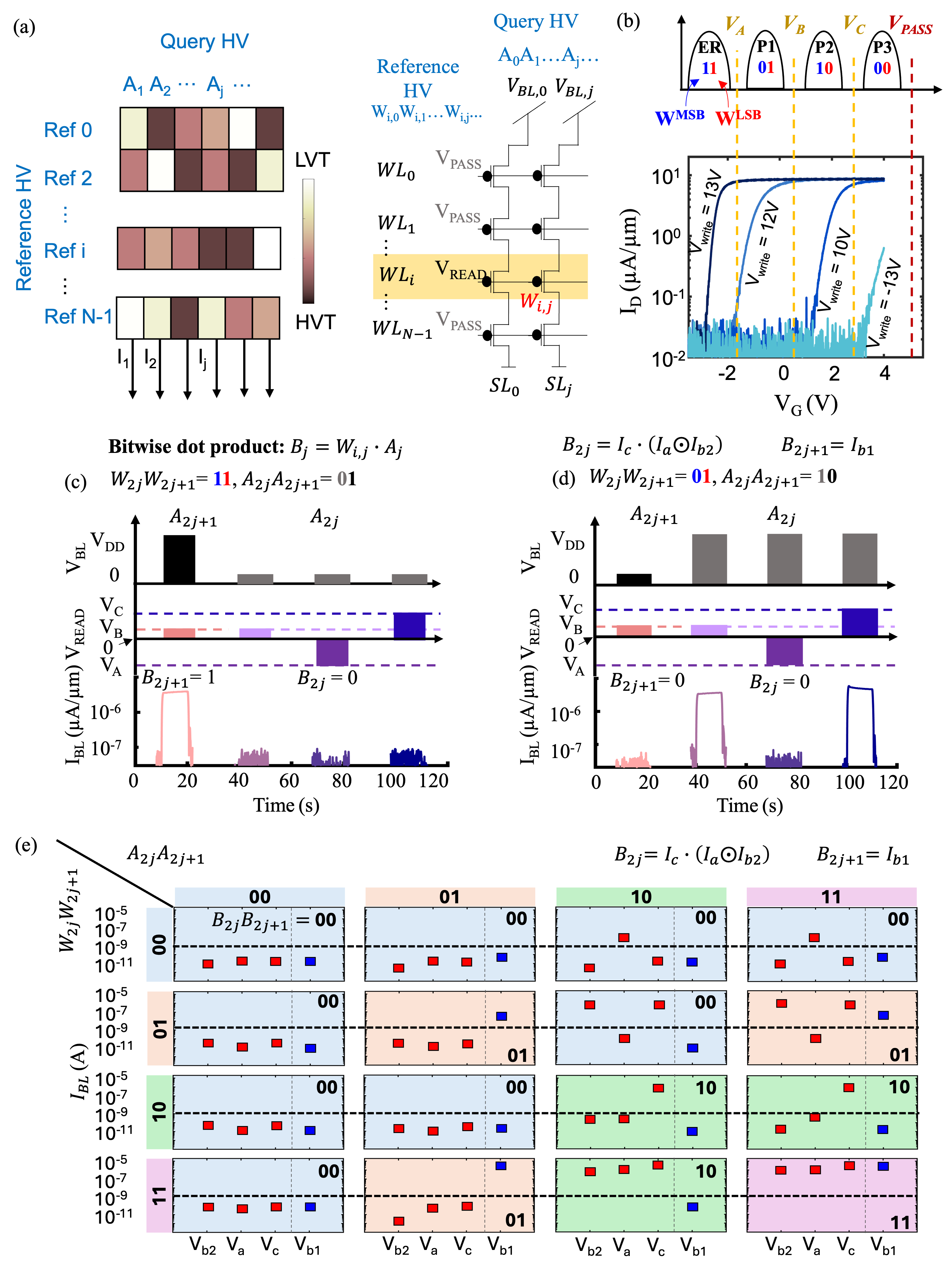}
\vspace{-0.2in}
\caption{(a) The reference HVs are stored along the WLs while the query HVs are passed as BL voltages. Bitwise MLC dot product between the query HV passed at the bitline and the stored reference HVs is demonstrated. (b) The two-bit output is read using a 4-step current read process at different read reference voltages, first at $V_B$, to obtain the $B_{2j+1}$ of the output and then at $V_A$, $V_B$ and $V_C$ to obtain the $B_{2j}$. The transient current outputs for two different stored values and inputs are shown in (c) and (d)\textcolor{black}{; time axis is in $\mu$s}. The truth table for bitwise MLC dot product is demonstrated with the output current for the 4-step read process shown for each combination. (e) The output is computed by digitizing the current. The current are measured using a DC spot read. The black dashed line is the threshold for digitizing the current. }
\label{array}
\end{figure*}

Digitized outputs are stored in a local buffer, and additional logic block completes the dot product by summing each binary outputs at the periphery of the NAND array. The supporting circuitry used for digitizing bitline currents and processing results is shown in Fig.\ref{intro}(d). While the current demonstration focuses on MLC operation, the method can be extended to higher logic levels, enabling multi-bit HD computing in 3D-NAND. \textcolor{black}{The number of read sensing steps per cell scales as $2^n$ for $n$-bit operation (1 for SLC, 4 for MLC, 16 for QLC), increasing the per-tile read latency accordingly (Supplementary Fig.~S8). However, since $n$-bit storage increases the data capacity per tile by $n\times$, the number of tiles required to store a fixed-size database is reduced by the same factor. The net total search time therefore scales as $T_{\text{total}} \propto 2^n / n$, increasing only $\sim$2$\times$ from SLC to QLC for the same database---far less than the $16\times$ per-cell cost increase.}

\section{\textcolor{black}{\textit{In-situ} Hamming similarity computation in FE-NAND}}

\begin{figure*}[!b]
\centering
\includegraphics[width = \textwidth]{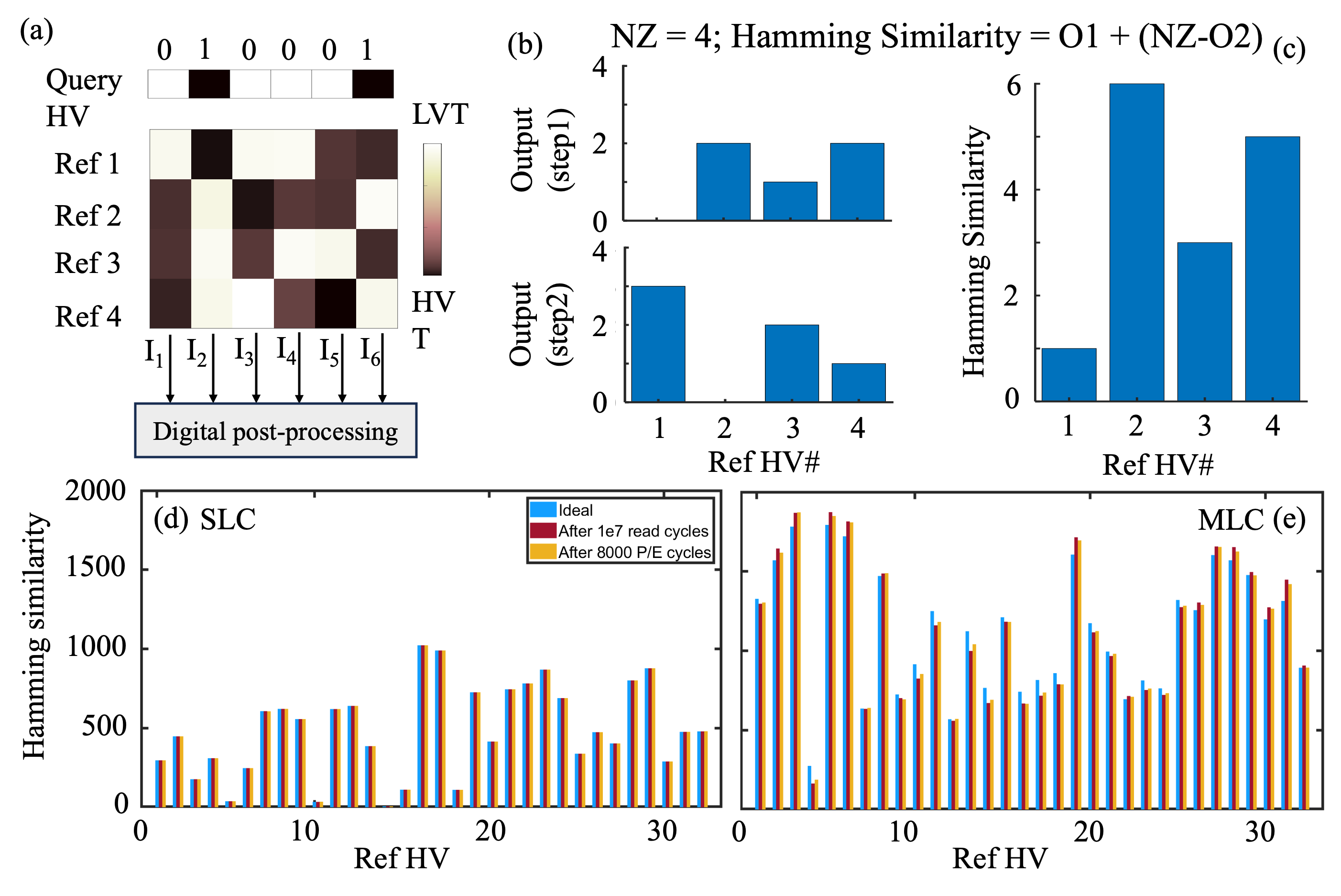}
\vspace{-0.2in}
\caption{(a) Schematics showing different $V_T$ states of the cells in an array with four 6-\textit{dim} hypervectors stored in them. Reference HV is stored in the NAND array. Query HV is given as input in the BLs of the array. (b) Summed dot product as a function of different Ref HVs for the two-step process to calculate the Hamming similarity. (c) Hamming similarity vs different Ref HVs. The Hamming similarity is highest in the Ref HV2. Therefore, Query HV matches the best with reference HV2. Hamming similarity for 32 HVs stored in an ideal array (blue) and the array after 10\textsuperscript{7} read cycles (brown) and 8000 P/E cycles (yellow) shows minimal effect to (d) SLC and (e) MLC operation. }
\label{hamming}
\end{figure*}

Building on the demonstrated \textit{bitwise} MLC dot product capability, we now show how FE-NAND arrays can support complete database search by enabling \textit{in-situ} Hamming similarity computation. This operation forms the core of associative search in HD computing and is central to applications such as Open Modification Search (OMS) in proteomics. \textcolor{black}{To demonstrate array-level Hamming similarity computation, we employ physics-based SPICE simulations in which each FeFET is modeled using the Preisach-based ferroelectric switching model and Virtual Source FET model, both calibrated against measured $I_D$--$V_G$ data from fabricated devices (Supplementary Note~6).} As shown in Fig.\ref{hamming}(a), four six-dimensional reference hypervectors (Ref HVs) are programmed into a \textcolor{black}{simulated} $4 \times 6$ FE-NAND array, with one hypervector per row (wordline). The query hypervector (Query HV) is passed at the bitlines, and the \textit{bitwise} dot products between the query and each reference HV are computed by digitizing the bitline currents at the corresponding sense lines (SLs). These dot product values are then summed  column-wise in digital domain to yield the total similarity score. Fig.\ref{hamming}(b) presents the summed dot products for both $Q \cdot R$ (O1) and $\overline{Q} \cdot R$ (O2) across all four reference HVs. The final Hamming similarity values, computed externally using the outputs from the array as O1 + (NZ - O2), where NZ is the number of zeros in the query hypervector, are shown in Fig.\ref{hamming}(c). Among the four references, HV2 exhibits the highest Hamming similarity, indicating the closest match with the input query HV. This demonstrates the feasibility of accurate similarity search within the FE-NAND array.

To further assess robustness, we simulate large-scale Hamming similarity computation under device variability and read and write cycling conditions. Fig.\ref{hamming}(d) shows the simulated Hamming similarity for single-level cell (SLC) operation in a $32 \times 1024$ FE-NAND array after $10^7$ read cycles and 8000 program/erase (P/E) cycles, assuming a threshold voltage variation $\sigma_{V_T}$ of 200~mV, a nominal variation reported in FE-NAND devices\cite{yoon2023qlc}. The results show no observable degradation in similarity values, with the correct reference HV consistently selected—demonstrating the intrinsic error resilience of HD computing. Additionally, Fig.\ref{hamming}(e) presents simulated results for multi-level cell (MLC) operation under identical stress conditions. After $10^7$ read cycles, the similarity scores exhibit minor deviations from the ideal case; however, the highest similarity still corresponds to the correct reference HV (HV5), indicating reliable operation. After 8000 P/E cycles, the highest similarity shifts to HV3, deviating from the expected result. This suggests that while MLC operation remains reliable under read stress, P/E cycling can introduce errors beyond a certain endurance threshold—highlighting the trade-off between density and robustness in multi-level FE-NAND systems.  

\section{System-level evaluation: OMS benchmarking}

Having demonstrated \textit{in-situ} multi-level dot product and Hamming similarity operations within FE-NAND arrays, we now evaluate the effectiveness of this architecture in a practical, large-scale setting. As a representative workload, we consider Open Modification Search (OMS), a critical step in mass spectrometry–based proteomics, which involves high-throughput similarity matching across vast protein databases. This task places significant demands on memory capacity, energy efficiency, and search accuracy—making it an ideal benchmark for in-storage HD computing.

\begin{figure*}[!b]
\centering
\includegraphics[width = \textwidth]{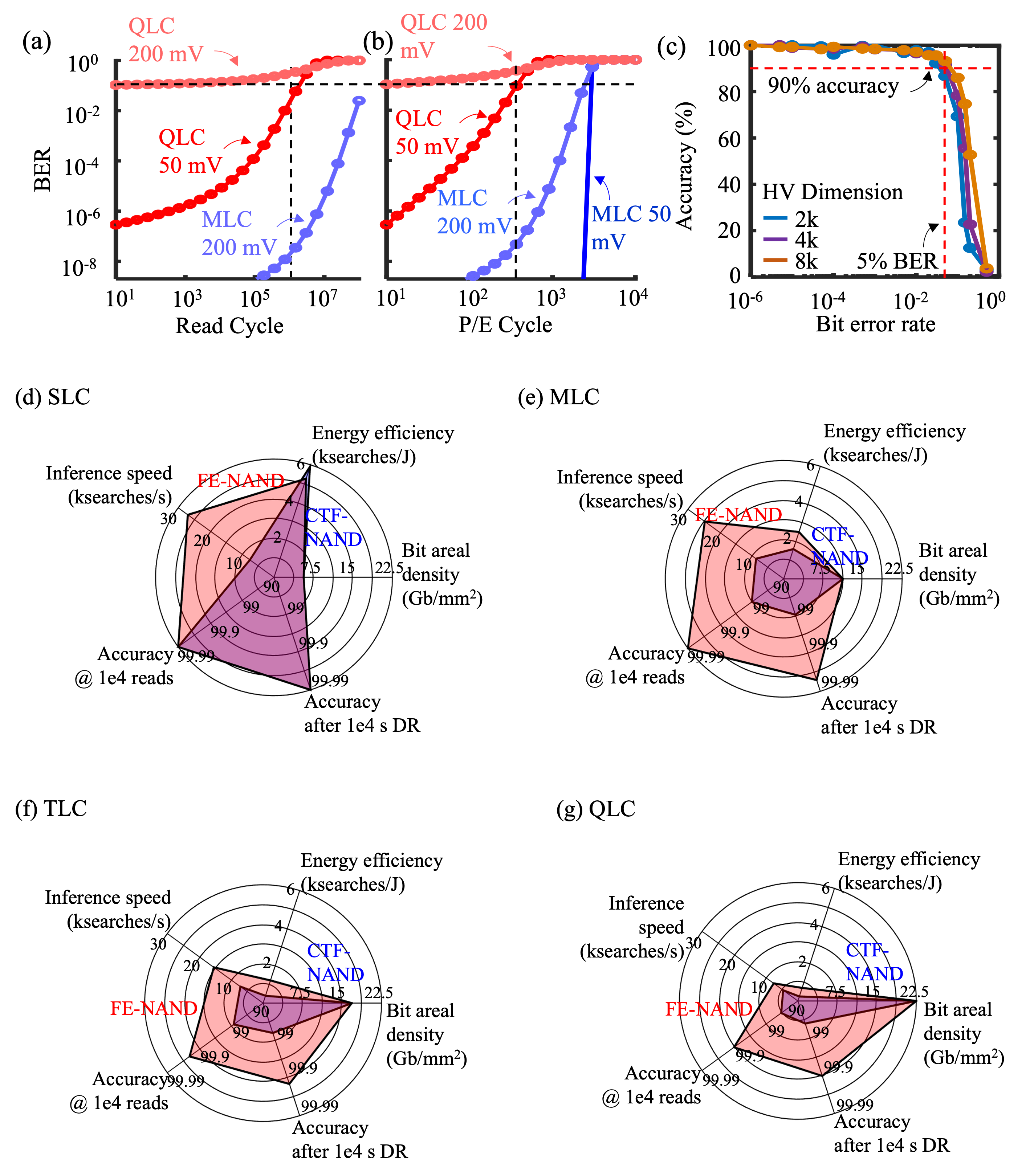}
\vspace{-0.2in}
\caption{The bit error rate for MLC and QLC operation with $\sigma_{V_T}$ of 50 and 200 mV (a) up to 10\textsuperscript{7} read cycles and (b) 8000 P/E cycles. (c) The error resilience of hyper-dimensional computing is shown with more than 90\% accuracy achieved for BER upto 5\%. (d-g) Benchmark showing inference speed, energy efficiency, bit area density and accuracy after 10k reads and 10\textsuperscript{4} s of data retention is shown for FE-NAND (red) and conventional CTF NAND (blue) for different logic scaling (SLC to QLC).\cite{cai2015data,cai2018read,goda2021recent,yang2025high,yang2025retention,papandreou2019characterization}}\label{benchmark}
\end{figure*}

Fig.\ref{benchmark}(a,b) presents the simulated bit error rate (BER) as a function of read cycles and program/erase (P/E) cycles for multi-level cell (MLC) and quad-level cell (QLC) operation, assuming threshold voltage variations ($\sigma V_T$) of 50 mV and 200 mV. While a device-to-device variation of 50~mV results in BERs below 5\% for up to $10^6$ read cycles, the 200 mV variation expected in scaled nodes leads to significantly higher BERs, particularly at higher logic levels such as QLC. These simulations highlight the need for error-tolerant architectures in enabling reliable high-density memory operation under realistic process variation.

To assess the impact of BER on application-level accuracy, we simulate OMS performance by evaluating the iPRG2012 dataset (15,867 spectra) against a reference library combining the yeast and human HCD spectral libraries (1,162,392 spectra)\cite{MassIVE2024,kang2022massively,chalkley2014iprg,selevsek2015reproducible}. Due to the inherent robustness of HD computing, the platform retains over 90\% accuracy even at BERs as high as 5\%. Furthermore, increasing hypervector dimensionality (i.e., number of bitlines) improves error tolerance, as shown in Fig.\ref{benchmark}(c). In the pristine state, the FE-NAND platform achieves over 99.99\% accuracy—comparable to current state-of-the-art systems—while offering substantially better energy efficiency and inference speed.  

We also benchmark FE-NAND against conventional charge-trap flash (CTF) NAND across different logic levels, evaluating performance for 15,000 query vectors matched against a 1-billion-vector reference library. Fig.\ref{benchmark}(d–g) summarize the results using spider plots that compare inference speed, energy efficiency, and accuracy after $10^4$ reads and $10^4$ s of data retention at equivalent storage density. At higher logic levels, FE-NAND consistently outperforms CTF-NAND in energy efficiency and speed. For example, in SLC mode, FE-NAND offers a 2.5$\times$ speed improvement over CTF-NAND with comparable energy usage. While both platforms maintain similar accuracy at low logic levels, FE-NAND exhibits significantly better accuracy at higher logic levels due to its superior retention and disturb resilience. Specifically, after $10^4$ s of retention, FE-NAND maintains $>$99.9\% accuracy, enabled by $<$1\% retention loss, whereas CTF-NAND experiences a notable accuracy drop. Similarly, after $10^4$ read operations, FE-NAND sustains high accuracy due to the implemented refresh-based disturb mitigation scheme.

Table 1 compares our FE-NAND platform with alternative in-storage and near-memory compute solutions for OMS tasks, including HyperOMS on GPU\cite{kang2022massively}, RapidOMS on FPGA\cite{pinge2024rapidoms}, and in-memory compute accelerators based on DRAM\cite{kang2024dram}, ReRAM\cite{fan2024efficient}, and 3D NAND\cite{hsu2023storage}. For large-scale datasets and high-dimensional hypervectors, in-storage compute using FE-NAND demonstrates over 700$\times$ lower latency and more than 22,000$\times$ improvement in energy efficiency compared to GPU- and near-memory-based approaches, all while maintaining comparable accuracy.

\textcolor{black}{The performance advantage of FE-NAND arises from two synergistic factors. At the device level, the QLC-compatible memory window (7.5~V) and low operating voltages ($<$14~V) maximize bit-area density while minimizing per-operation energy relative to CTF NAND ($>$20~V). At the architecture level, in-situ processing eliminates the storage-to-compute data movement that dominates energy and latency in GPU-, FPGA-, and DRAM-based solutions, while leveraging the inherent page-level parallelism of 3D NAND (thousands of bitlines sensed per page read) to scale throughput with array size.}

\textcolor{black}{Unlike NOR-type FeFET CAM or crossbar architectures that achieve $O(1)$ parallel search, FE-NAND processes one wordline per sensing cycle ($O(N)$ complexity). However, each cycle reads an entire page (thousands of bitlines in parallel), and the ultra-high storage density ($>$100~Gb/mm$^2$) enables the full TB-scale database to reside in-storage. For workloads where data movement dominates total cost---as is the case for the TB-scale proteomics database considered here---FE-NAND's density advantage more than compensates for the sequential row access, achieving $>$700$\times$ lower latency and $\sim$28$\times$ better energy efficiency than the best-reported crossbar solution (Table~1). While analog crossbar arrays enable single-step dot products, their on-chip capacity is limited to $\sim$MB-scale, requiring repeated off-chip data fetches for TB-scale workloads that dominate total system energy. FE-NAND's $>$100~Gb/mm$^2$ projected density enables the full database to remain in-storage, and its digital multi-step readout avoids the analog noise, IR drop, and sneak-path issues that constrain large crossbar arrays, achieving $\sim$28$\times$ better total energy efficiency than the best-reported ReRAM crossbar solution (Table~1). A detailed comparison of the architectural trade-offs is provided in Supplementary Note~10.}




\section{Conclusion}

This work establishes FE-NAND as a scalable, in-situ compute platform for hyperdimensional (HD) search workloads, demonstrating multi-bit dot product and Hamming similarity directly within fabricated NAND arrays. By executing full HD pipelines entirely in-storage including multi-level readout, transient current digitization, and associative matching—we overcome the fundamental energy and bandwidth bottlenecks of von Neumann architectures. Our system-level benchmarks on Open Modification Search (OMS) workloads validate this approach, revealing close to 1000$\times$ throughput and over 10,000$\times$ energy efficiency gains compared to near-memory baselines.

The implications of these results extend across device, circuit, and system domains. At the device level, the integration of ferroelectrics into NAND strings enables dense, disturb-free multi-level states with $<$14~V operation and robust retention—key to supporting QLC-compatible logic with in-storage compute functionality. At the architecture level, FE-NAND allows pairing ultra-high areal density with compute capability, offering a route to $>$100~Gb/mm\textsuperscript{2} and $>$1000-layer stacks without sacrificing reliability or accuracy. Systemically, this paves the way for massive similarity search at sub-nJ/query—addressing urgent needs in proteomics, AI inference, and genomic alignment.

Looking forward, FE-NAND opens the door to a new class of in-storage intelligent accelerators that couple the maturity of NAND flash fabrication with non-von Neumann architectures. By embedding logic into standard memory backends, this platform could enable edge-native analytics, real-time database search, and privacy-preserving AI inference within the footprint and cost of commodity storage. 


\newpage
\begin{spacing}{1}
\newpage

\newpage

\newpage

\newpage
\end{spacing}

\begin{landscape}
\begin{table}
\centering
\resizebox{\linewidth}{!}{\setlength{\tabcolsep}{10pt} 
\renewcommand{\arraystretch}{1.5} 
\begin{tabular}{ccccccc}
\hline
\textbf{Architecture}                                                          & \textbf{\begin{tabular}[c]{@{}c@{}}GPU\cite{kang2023accelerating,kang2022massively}\\ \end{tabular}} & \textbf{\begin{tabular}[c]{@{}c@{}}FPGA\cite{chalkley2014iprg}\\ \end{tabular}}             & \textbf{\begin{tabular}[c]{@{}c@{}}DRAM\cite{kang2024dram}\\ \end{tabular}}         & \textbf{\begin{tabular}[c]{@{}c@{}}MLC ReRAM\cite{fan2024efficient}\end{tabular}} & \textbf{\begin{tabular}[c]{@{}c@{}}3D NAND\cite{hsu2023storage}\end{tabular}} & \textbf{FeNAND (this work)}                                            \\ \hline
\textbf{Compute type}                                                          & -                                                                    & -                                                                                 & Near-memory                                                                   & Near-memory                                                               & In-memory                                                                & In-memory                                                              \\ \hline
\textbf{Algorithm}                                                             & HOMS-TC                                                              & RapidOMS                                                                          & HyperOMS-PIM-DRAM                                                             & HyperOMS-PIM-ReRAM                                                        & HyperOMS-3DNAND                                                          & HyperOMS-FeNAND                                                        \\ \hline
\textbf{Technology Node}                                                       & \begin{tabular}[c]{@{}c@{}}RTX 4090\\ (5nm)\end{tabular}             & \begin{tabular}[c]{@{}c@{}}Versal Prime VM1802 (7 nm) \\ SmartSSD 2.0\end{tabular} & \begin{tabular}[c]{@{}c@{}}22 nm DRAM (DDR4) \\ 28 nm node-compute\end{tabular} & 40 nm RRAM with 3M cells                                                  & \begin{tabular}[c]{@{}c@{}}NAND: 14 nm\\ ASIC: 7 nm FinFET\end{tabular}    & \begin{tabular}[c]{@{}c@{}}FeNAND: 14 nm\\ ASIC:7 nm FinFET\end{tabular} \\ \hline
\textbf{Inference Speed}                                                                 & 1$\times$ (23 min)                                                           & 1.1$\times$                                         & 2.43$\times$                                                                         & 5.42$\times$                                                                      & 423$\times$                                                                     & 737$\times$                                                                  \\ \hline
\textbf{Inference Energy Efficiency}                                                     & 1$\times$ (454 kJ)                                                            & 11.0625$\times$                                     & 101$\times$                                                                          & 7725$\times$                                                                      & 7230$\times$                                                                    & 22146$\times$                                                                  \\ \hline
\textbf{Total Energy Efficiency}                                                     & 1$\times$ (454 kJ)                                                            &  9.24$\times$                                     & 36$\times$                                                                         & 502$\times$                                                                      & 4683$\times$                                                                    & 14277$\times$                                                                  \\ \hline
\textbf{\begin{tabular}[c]{@{}c@{}}Capacity Limit\\ (per module)\end{tabular}} & 24 GB                                                                 & 32 GB                                         & 128 GB                                                                         & 3M cells                                                                  & 16 TB                                                                     & \textgreater{}100 TB                                                    \\ \hline
\textbf{Data density}                                                          & -                                                                    & -                                                                                 & -                                                                             & -                                                                         & 20-30 Gb/mm\textsuperscript{2}                                                              & \textgreater{}100 Gb/mm\textsuperscript{2}                                               \\ \hline
\end{tabular}%
}
\caption{As a benchmarking standard, we assume a repository with one billion reference HVs and 15k query HVs. FE-NAND devices retain the advantages arising from the parallelism and read energy efficiency of CTF NAND while achieving significantly higher data densities. The inference energy for GPU is inclusive of the write energy as the write and read energies cannot be separated.}
\label{tablebm}
\end{table}
\end{landscape}
\newpage

\input{addendum.tex}

 \bibliography{references}
 \input{methods}

\end{document}


\maketitle
\vspace{-.3in}
\begin{affiliations}
\item School of Electrical and Computer Engineering, Georgia Institute of Technology, Atlanta, GA, USA.

\item School of Electrical and Computer Engineering, University of Texas Dallas, 
Richardson, TX, USA.

\item School of Electrical and Computer Engineering, University of California San Diego, San Diego, CA, USA.

\item Institute of Materials, Georgia Institute of Technology, Atlanta, GA, USA.

\item Institute of Matter and Systems, Georgia Institute of Technology, Atlanta, GA, USA.

\item Advanced Device Research Laboratory, Semiconductor Research Center, Samsung Electronics Co., Ltd., South Korea.

\item School of Materials Science and Engineering Engineering, Georgia Institute of Technology, Atlanta, GA, USA.

\item[\textsuperscript{*}] Corresponding authors. Email: pravindran6@gatech.edu, akhan40@gatech.edu.  

\end{affiliations}

\newpage
\section*{Contents}

\textbf{Supplementary note 1. Fabrication process for FE-NAND arrays\\
Supplementary note 2. Reliability characterization of FEFET with 8/3/8 gate stack\\
Supplementary note 3. Benchmark of device characteristics\\
Supplementary note 4. Experimental setup for array characterization\\
Supplementary note 5. Demonstration of SLC dot product\\
Supplementary note 6. Physics-based model of FE-NAND\\
Supplementary note 7. Bit-error rate calculations\\
Supplementary note 8. System level benchmarking\\
Supplementary note 9. Multi-bit operation\\
\textcolor{black}{Supplementary note 10. Architectural comparison}}

\newpage
\section*{Supplementary note 1. Fabrication process for FE-NAND arrays}

The below figure shows the fabrication process flow for FE-NAND arrays. The source and drain regions are implemented through ion implantation (BF\textsubscript{2}, 15keV). After dopant activation, the screening oxide is etched and silicon MESA is patterned. Following this, the gate stacks are deposited through thermal atomic layer deposition (THALD) of HZO and Al\textsubscript{2}O\textsubscript{3}. Tungsten gate electrodes are sputtered and rapid thermal annealing (RTA) is performed to crystallize the HZO layers. RTA is done at 500\textsuperscript{$\circ$}C for 30 seconds. The gate electrodes are patterned by dry etching the W layer using RIE. The device is completed with a standard interlayer dielectric (ILD) and metallization process. A forming gas anneal at 400\textsuperscript{$\circ$}C for 30 min is performed to improve the quality of the IL.

\begin{figure}[!h]
    \centering
    \includegraphics[width=0.5\linewidth]{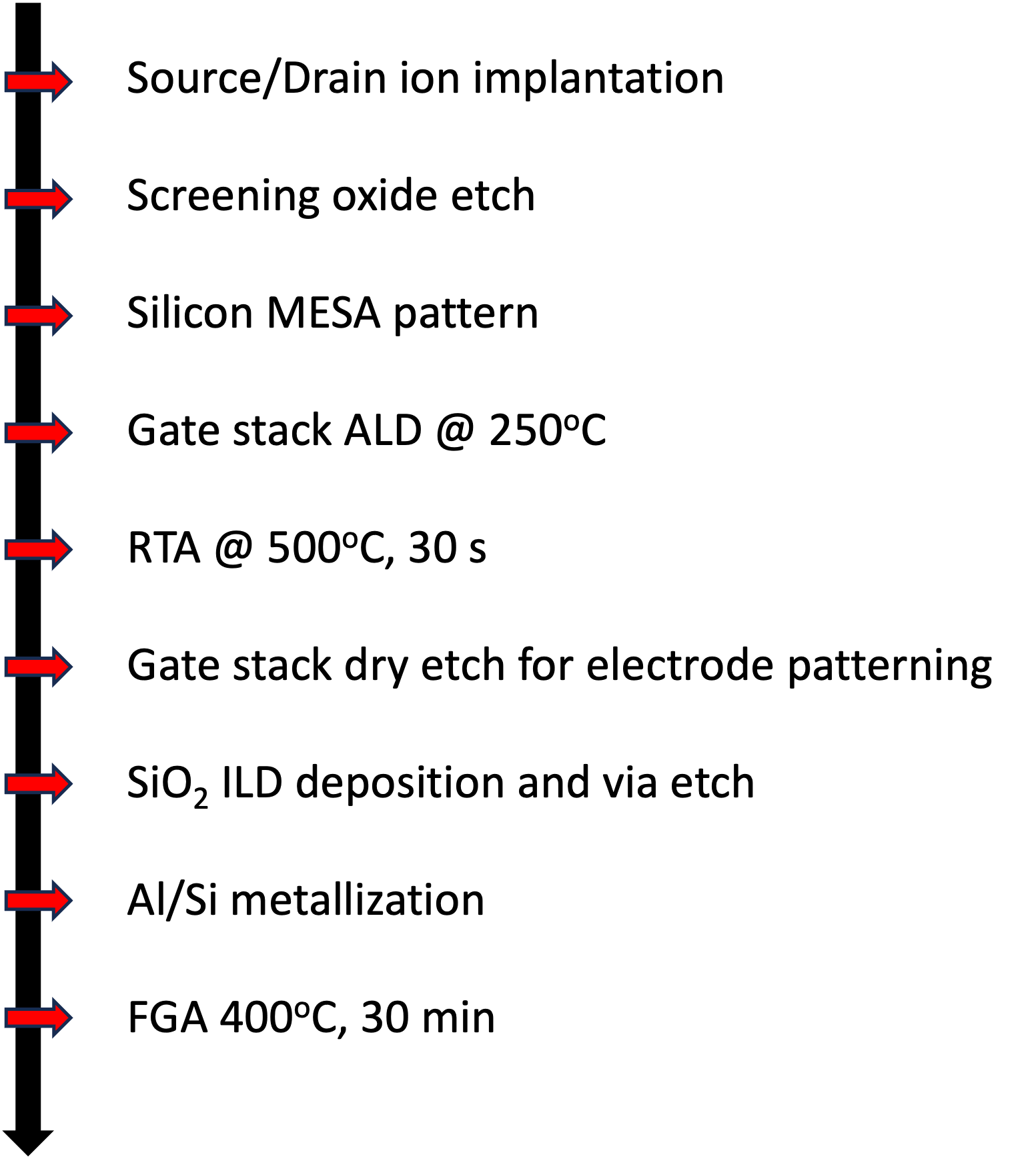}
    \caption{Process steps for fabricating FE-NAND arrays.}
    \label{S1}
\end{figure}
\newpage

\newpage
\section*{Supplementary note 2. Reliability characterization of FEFET with 8/3/8 gate stack}

Disturb, retention and endurance are the three important reliability metrics for NAND applications. 

\noindent \textbf{(1) Disturb: } While accessing a cell, a pass voltage is used to turn on the other unaddressed cells along the bitline in a NAND array. Earlier reports have shown that electron trapping causes a positive shift in the threshold voltage in the program states. We adopt the disturb mitigation scheme presented in \cite{venkatesan2024disturb} to reduce disturb in the different program states. The disturb characteristics with the disturb mitigation scheme where refresh pulses are applied every 10 disturb cycles is shown in Fig. S2(a). A pass voltage of 6 V is used for these measurements along with a refresh pulse of -10 V to detrap the trapped electrons. 

\noindent \textbf{(2) Endurance: } Bipolar cycling is performed to obtain the endurance of the gate stack as shown in Fig. S2(b). At 10 V cycling with 10 $\mu$s pulses shows an endurance of 10000 P/E cycles where the MW closes. Importantly, the MW stays steady upto 3e3 cycles before degrading rapidly.

\noindent \textbf{(3) Retention: } The retention characteristics are obtained by measuring the transfer characteristics of different states over time. We see that the MW stays fairly constant after 1e4 s. A detailed account on the origin of the stable MW in TDL gate stack based FEFETs is presented in \cite{venkatesan2025enhanced}. 

\begin{figure}[!h]
    \centering
    \includegraphics[width=0.95\linewidth]{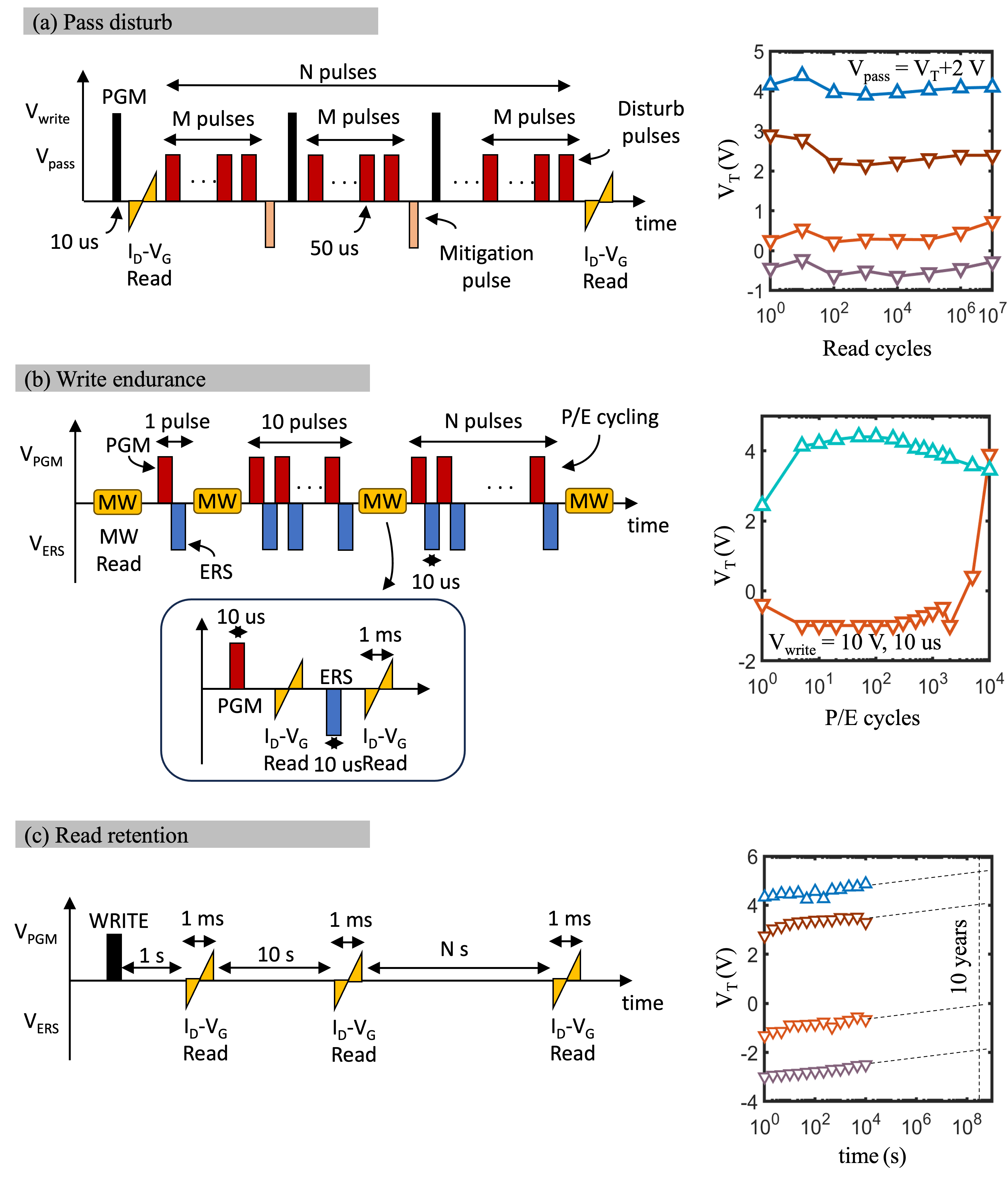}
    \caption{(a) Multi-level pass disturb shows minimum disturb with the application of periodic refresh pulses. (b) P/E endurance shows steady MW up to 3e3 cycles before undergoing rapid degradation. (c) Robust retention is also demonstrated up to 1e4 s. }
    \label{S1}
\end{figure}

\noindent \textbf{(4) Retention comparison between gate-blocking layer and tunnel-dielectric layer FEFETs: } Multi-temperature retention characterization of the tunnel dielectric layer (TDL) FEFETs and gate blocking layer (GBL) FEFETs reveal that the initial MW in GBL FEFETs decreases with increasing temperature while the that in TDL FEFETs remains constant even at higher temperatures. Further, the MW loss in GBL FEFETs remains constant as a function of temperature while the MW loss in TDL FEFETs is very low at RT and increases at higher temperatures \cite{venkatesan2024demonstration}. The time and temperature dependence evolution of MW in FEFETs can be modeled using a time-dependent Arrhenius model\cite{venkatesan2025enhanced}, where, 

\begin{align*}
    MW(t,T) &= MW(0,T) - \Delta MW(t,T)\\
    \Delta MW(t,T) &= Ae^{-E_a/kT}t^n\\
    MW(0,T) &= MW(0,0)\cdot e^{-E_{a,0}/kT}
\end{align*}

\begin{figure}[!t]
    \centering
    \includegraphics[width=0.9\linewidth]{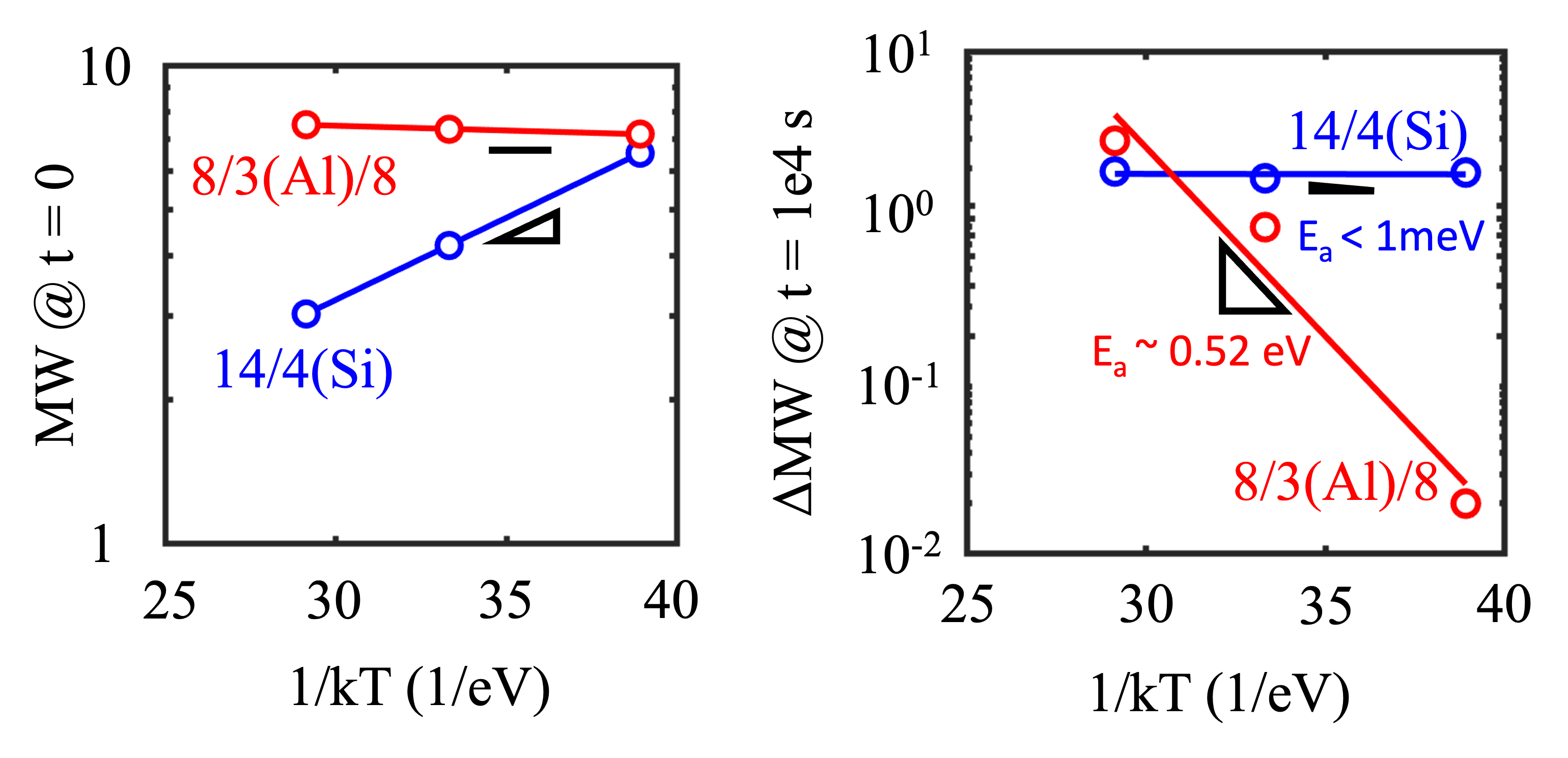}
    \caption{(a) Initial MW in gate blocking layer FEFETs decreases with increasing temperature while that in tunnel dielectric layer FEFETs remains constant across temperatures. (b) The GBL FEFETs exhibits a constant MW loss across temperatures while TDL show increased MW loss with increasing temperature. }
    \label{S1}
\end{figure}

\newpage
\begin{landscape}

\begin{table}
\centering
\begin{minipage}{1\columnwidth}    
\section*{Supplementary note 3. Benchmark of device characteristics}
\vspace{1in}
\resizebox{1\columnwidth}{!}{%
\begin{tabular}{cccccccc}
\hline
\textbf{}                    & \textbf{CTF}   & \textbf{\begin{tabular}[c]{@{}c@{}}Yoon et al., \\ VLSI 2023 \cite{yoon2023qlc}\end{tabular}} & \textbf{\begin{tabular}[c]{@{}c@{}}Lim et al., \\ IEDM 2023 \cite{lim2023comprehensive}\end{tabular}} & \textbf{\begin{tabular}[c]{@{}c@{}}Yoo et al., \\ VLSI 2024 \cite{yoo2024highly}\end{tabular}} & \textbf{\begin{tabular}[c]{@{}c@{}}Kuk et al., \\ IEDM 2024 \cite{kuk2024superior}\end{tabular}} & \textbf{\begin{tabular}[c]{@{}c@{}}Joh et al., \\ IEDM 2024 \cite{joh2024oxide}\end{tabular}}                        & \textbf{This work}                                    \\ \hline
\textbf{Semiconductor}       & Poly Si        & Poly Si                                                                    & Poly Si                                                                           & IGZO                                                                               & Si                                                                                 & IZO                                                                                              & SOI                                                   \\ \hline
\textbf{FE}                  & -              & HZO                                                                        & \begin{tabular}[c]{@{}c@{}}HZO\\ (12.5 nm)\end{tabular}                           & \begin{tabular}[c]{@{}c@{}}HZO\\ (18 nm)\end{tabular}                              & \begin{tabular}[c]{@{}c@{}}HZO\\ (8.6 nm)\end{tabular}                             & \begin{tabular}[c]{@{}c@{}}HZO\\ (9 nm)\end{tabular}                                             & \begin{tabular}[c]{@{}c@{}}HZO\\ (16 nm)\end{tabular} \\ \hline
\textbf{DE material}         & -              &                                                                            & SiO\textsubscript{2} (3 nm)                                                                       & Al\textsubscript{2}O\textsubscript{3} (3 nm)                                                                       & \begin{tabular}[c]{@{}c@{}}HfO\textsubscript{2} (3.2 nm)/ \\SiO\textsubscript{2} (3 nm)\end{tabular}                                                         & \begin{tabular}[c]{@{}c@{}}Si\textsubscript{3}N\textsubscript{4} (4 nm)/ \\SiO\textsubscript{2}(1.5 nm)/ \\ Si\textsubscript{3}N\textsubscript{4}(2 nm)/ \\SiO\textsubscript{2}(2.5 nm)\end{tabular} & Al\textsubscript{2}O\textsubscript{3} (3 nm)                                          \\ \hline
\textbf{MW (V)}              & 7.5 (QLC)      & 10.5                                                                       & 4.6                                                                               & 17.8                                                                               & 12.2                                                                               & 14                                                                                               & 7.5                                                   \\ \hline
\textbf{Write voltage (V)}   & 20-30          & 15                                                                         & 17                                                                                & 15                                                                                 & 15                                                                                 & 20                                                                                               & 13.5                                                  \\ \hline
\textbf{Write endurance}     & 1e5-1e6        & 3e3                                                                        & 1e6                                                                               & 1e4                                                                                & 5e3                                                                                & \textless{}1e2                                                                                   & 1e4                                                   \\ \hline
\textbf{Pass disturb (\%)}   & \textless{}3\% & -                                                                          & 20\%                                                                              & -                                                                                  & -                                                                                  & \begin{tabular}[c]{@{}c@{}}1\% (after 1e5 \\ read cycles)\end{tabular}                           & $\sim$4\%                                             \\ \hline
\textbf{Retention loss (\%)} & \textless{}3\% & -                                                                          & 46\%                                                                              & 37\% after 1e3 s                                                                   & \textless{}1\%                                                                     & \textless{}5\%                                                                             & \textless{}1\%                                        \\ \hline
\end{tabular}%
}
\caption{Device level benchmarking showing QLC compatible memory window at low write voltage($<$ 14 V), record high retention along with pass disturb characteristics on par with CTF NAND.}
\label{tab:my-table}
\end{minipage}
\end{table}
\end{landscape}
\newpage

\section*{Supplementary note 4. Experimental setup for array characterization}

The experimental setup used to characterize the FE-NAND strings are shown in Fig. S4. The pass voltage is set at 6 V. The different WLs are accessed by using a switch matrix. The reference HVs are stored in WLs while the query HV is passed at the bitline. A 1 V pulse is used for a high input at the bitline while 0 V is used for a low input. The bitline currents are measured using a Keysight B1500. The currents are then post-processed digitally. \textcolor{black}{A current compliance of 100~$\mu$A was applied during DC spot read measurements on the Keysight B1500.}

\begin{figure}[!h]
    \centering
    \includegraphics[width=0.99\linewidth]{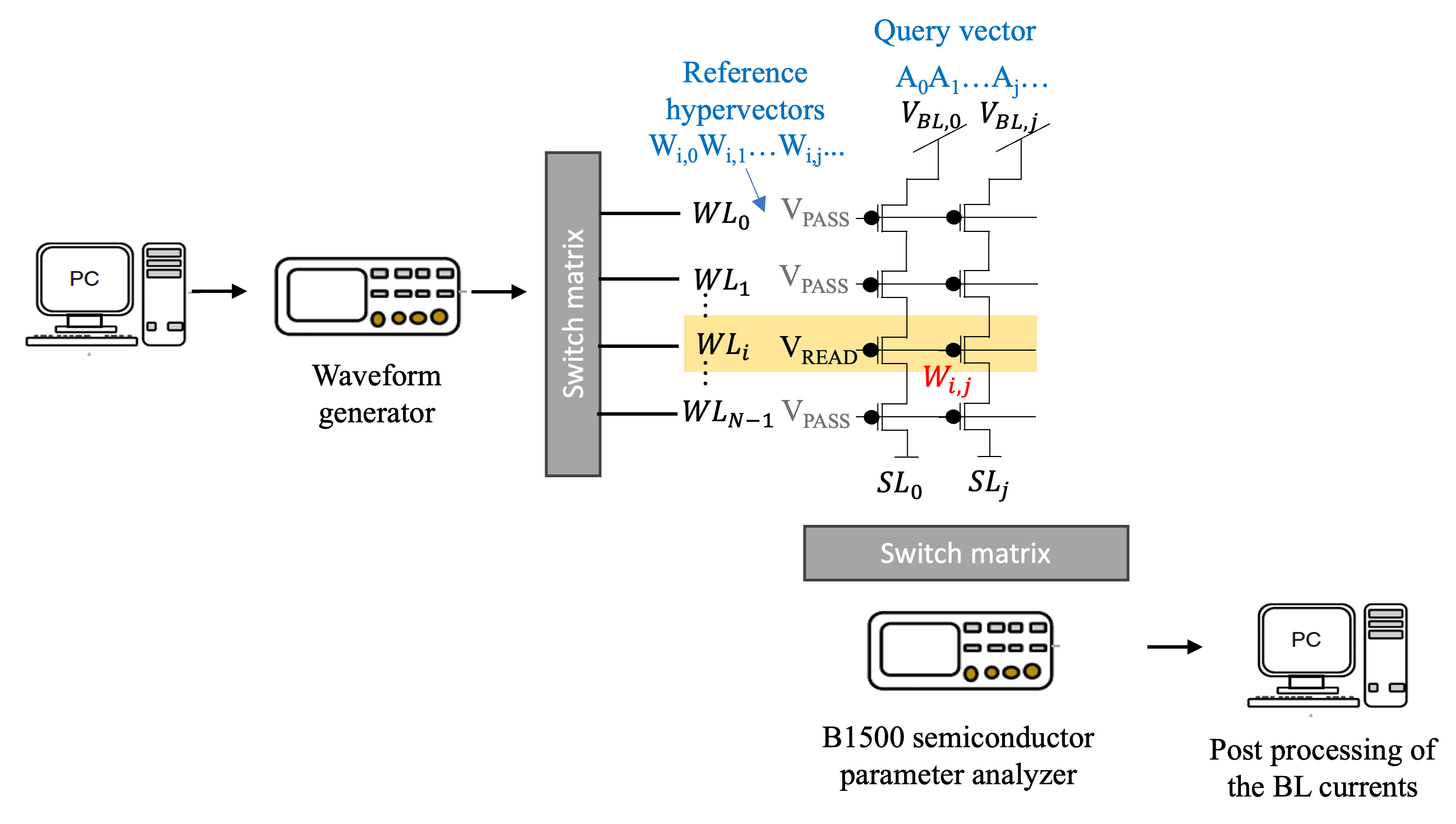}
    \caption{Experimental setup for electrical characterization of ferroelectric NAND strings. }
    \label{S2}
\end{figure}
\newpage

\section*{Supplementary note 5. Demonstration of SLC dot product}

SLC dot products are demonstrated by storing the reference bit in the cell and passing the query bit at the bitline. When a particular WL is addressed, the other WLs are passed using a 6 V pulse. The bitline current output is readout at a read voltage of V\textsubscript{G} = 1 V. A distribution of 100 dot products for all the combinations of reference and query bits are shown in Fig. S5(b). The output is digitized in post-processing assuming a current threshold of 1 nA. \textcolor{red}{These 100 repeated measurements per input--weight combination are performed experimentally on fabricated FE-NAND strings and confirm the reliability of the \textit{in-situ} dot-product compute primitive at the string level. The array-level Hamming similarity (Fig.~4) and system-level benchmarks (Fig.~5, Table~1) are subsequently obtained from experimentally calibrated physics-based simulations (Supplementary Note~6); experimental demonstration of a parallel multi-bitline array performing simultaneous \textit{in-situ} search remains an important direction for future work.}

\begin{figure}[!h]
    \centering
    \includegraphics[width=0.9\linewidth]{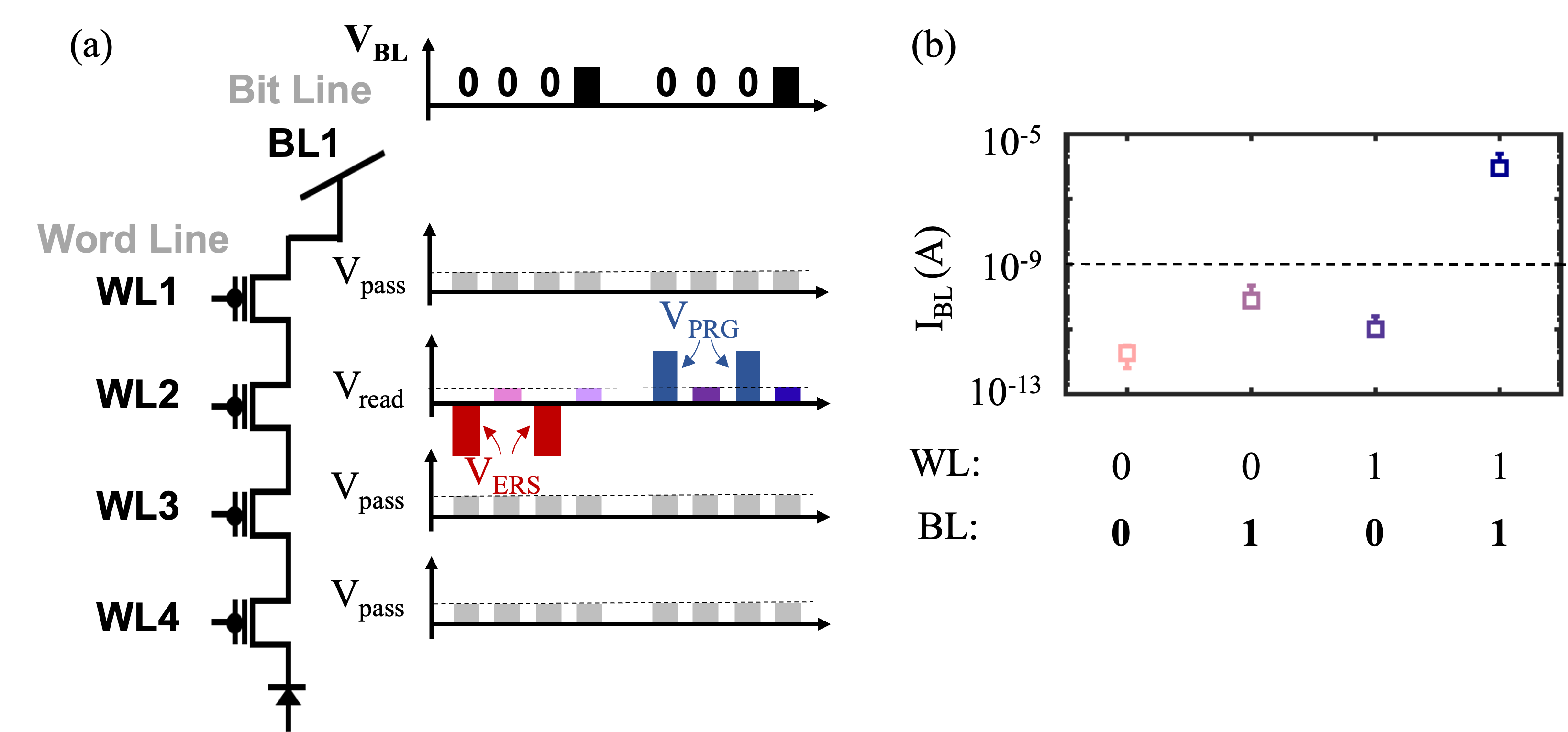}
    \caption{(a) Experimental demonstration of SLC dot product on a FE-NAND string for different WL states and BL inputs. (b) The distribution of bit lines currents for 100 dot products is shown. }
    \label{S2}
\end{figure}
\newpage

\section*{Supplementary note 6. Physics-based model of FE-NAND}

\begin{figure}[!h]
    \centering
    \includegraphics[width=0.9\linewidth]{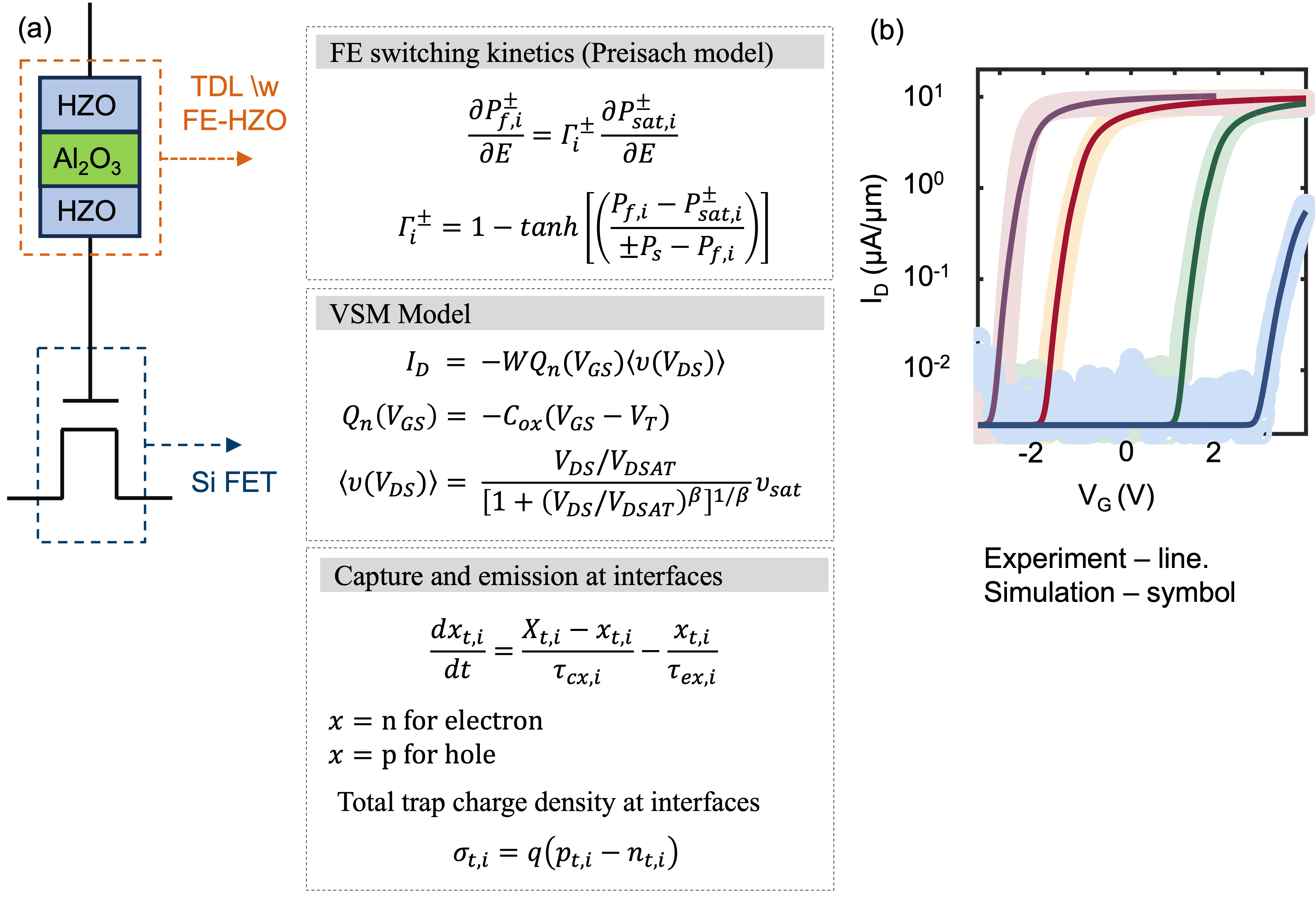}
    \caption{(a) Modeling trapping dynamics and switching kinetics of the TDL FE and virtual source model equations for the Si FET. (b) Calibration of the model with experimental $I_D-V_G$ data.  }
    \label{S2}
\end{figure}

Fig. S6(a) illustrates the modeling framework of a ferroelectric field-effect transistor (FeFET) incorporating a ferroelectric stack based on HZO with an interfacial Al\textsubscript{2}O\textsubscript{3} dielectric, integrated on top of a silicon FET. Ferroelectric polarization switching is described using a Preisach-based kinetic model, where the polarization change is governed by a field-dependent switching factor that captures the hysteretic nature of the ferroelectric material. The underlying FET operation is modeled using the Virtual Source Model (VSM). Additionally, the dynamics of charge trapping and de-trapping at interfaces are modeled through rate equations, incorporating carrier-type dependent capture and emission times. The total interfacial charge is then computed based on the trap occupancy, which plays a crucial role in modulating the threshold voltage and overall device hysteresis. Fig. S6(b) shows the calibration of the model with experimental I\textsubscript{D} – V\textsubscript{G} data.

\textcolor{black}{The NAND string SPICE model explicitly includes all unselected cells biased at $V_{pass} = 5$~V. The on-resistance of pass cells in the linear regime is much smaller than the subthreshold resistance of the selected cell. Furthermore, the experimental single-cell dot product measurements are performed on fabricated NAND strings where these effects are inherently present. The model is used for all array-level Hamming similarity simulations (Fig.~4) and large-scale BER calculations (Fig.~5).}

\vspace{0.1in}
\begin{table*}[!htbp]
\centering
\begin{tabular}{ccc}
\hline
\textbf{Parameter} & \textbf{Value} & \textbf{Description} \\
\hline
type      & 1                 & type = 1 for NMOS, type = -1 for PMOS \\\hline
Width     & 40e-6             & Width [m] \\\hline
etov      & 0e-7              & Equivalent thickness of dielectric at S/D-G overlap [cm] \\\hline
C\textsubscript{g,VT1}   & 6e-6              & Gate-to-channel areal capacitance at the virtual source [F/cm$^2$] \\\hline
C\textsubscript{g,VT2}   & 3.2e-6            & Gate-to-channel areal capacitance [F/cm$^2$] \\\hline
C\textsubscript{g,VT3}   & 4e-6              & Gate-to-channel areal capacitance [F/cm$^2$] \\\hline
C\textsubscript{g,VT4}   & 0.3e-6            & Gate-to-channel areal capacitance [F/cm$^2$] \\\hline
C\textsubscript{if}       & 1.2$\times$10$^{-13}$ & Inner fringing S or D capacitance [F/cm] \\\hline
C\textsubscript{of}       & 1.2$\times$10$^{-13}$ & Outer fringing S or D capacitance [F/cm] \\\hline
beta      & 1.8               & Saturation factor (nFET=1.8, pFET=1.6) \\\hline
alpha     & 3.5               & Charge Vt transition factor (typ. 3.0--4.0) \\\hline
R\textsubscript{s}        & 2200              & Access region resistance for s terminal [Ohms-$\mu$m] \\\hline
R\textsubscript{d}        & 2200              & Access region resistance for d terminal [Ohms-$\mu$m] \\\hline
DIBL      & 0.07              & Drain induced barrier lowering (DIBL) [V/V] \\\hline
n\textsubscript{0}        & 2                 & Subthreshold swing factor (typically 1.0--2.0) \\\hline
n\textsubscript{d}        & 0.4               & Punch-through factor [1/V] \\\hline
v\textsubscript{x,0}       & 1.5               & Virtual-source injection velocity [cm/s] \\\hline
$\mu$        & 150               & Low field mobility [cm$^2$/Vs] \\\hline
\end{tabular}
\caption{Device simulation parameters}
\end{table*}
\vspace{-0.1in}

\newpage



\section*{Supplementary note 7. Bit-error rate calculations}

The variation in $V_T$ is modeled using a gaussian distribution. The bit-error rate (BER) is computed using the below expression. The BER after different read and write cycles is calculated by shifting the distribution by the $\Delta V_T$ measured from Fig. S2. 

\begin{equation}
    BER = 1 - \int_{V_{REF,1}}^{V_{REF,2}} \mathcal{N}(V_{T,state} + \Delta V_{T},\,\sigma^{2})\
\end{equation}

\begin{figure}[!h]
    \centering
    \includegraphics[width=0.8\linewidth]{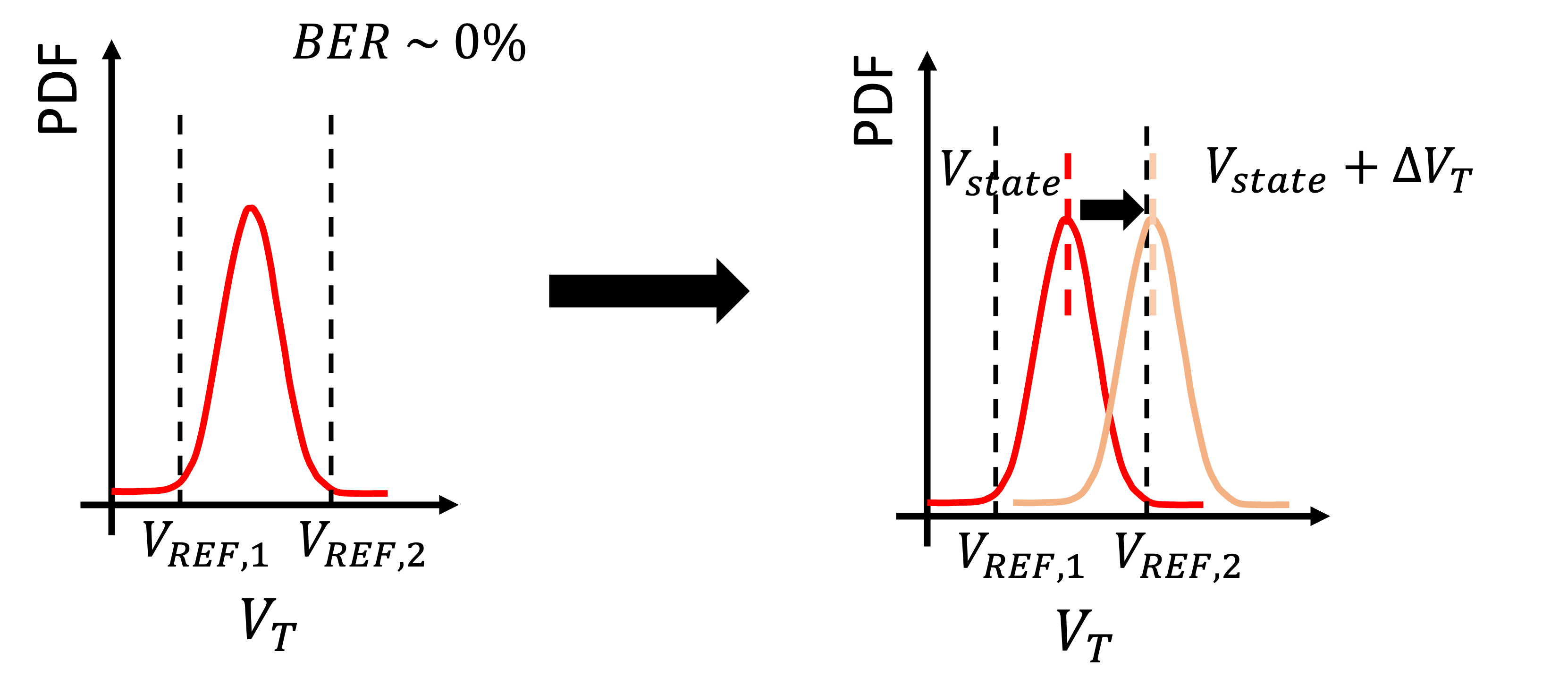}
    \caption{The bit-error rate (BER) is calculated by integrating between the reference voltages V\textsubscript{ref,1} and V\textsubscript{ref,2}. The evolution with different $V_T$ shifts is evaluated by shifting the distribution by the measured $\Delta V_T$. } 
    \label{S1}
\end{figure}
\newpage

\section*{Supplementary note 8. System-level benchmarking}

We assume the same number of WLs in both CTF and FE-NAND. We compute the energy efficiency and inference speed from energy and time shown below. They are computed using our benchmarking framework described in detail in the methods section. For benchmarking, we use 15,000 query inputs, and 1,000,000,000 pre-stored spectra as reference library.

\textcolor{black}{\noindent \textbf{Hierarchical modeling methodology:} System performance is modeled at four levels of abstraction: (1) \textit{Device level:} Experimentally calibrated Preisach + Virtual Source SPICE models (Supplementary Note~6) provide unit-cell FeFET parameters including $I_D$--$V_G$ characteristics, ferroelectric switching kinetics, and charge trapping dynamics. (2) \textit{Array level:} The NAND string model includes all wordlines with pass-cell parasitics; array-level RC parasitics (bitline resistance, wordline RC delay) are based on industry-grade 3D NAND chip-level parameters consistent with state-of-the-art process nodes at a 14~nm equivalent node. (3) \textit{Peripheral circuits:} Row decoders, sense amplifiers, page buffer latches, and ADCs are extracted from NeuroSim v1.5, which has been validated against silicon data for NAND-based architectures. The CMOS peripheral circuits are modeled at a 7~nm FinFET node using the ASAP7 PDK. (4) \textit{System level:} The in-storage compute architecture uses CUA (65~nm high-voltage) and CBA (7~nm low-voltage) heterogeneous 3D integration with Cu-Cu hybrid bonding at $\sim$1~$\mu$m inter-tier pitch. Energy and latency are computed using cycle-accurate simulation accounting for all data-movement, compute, and peripheral overhead.}

\textcolor{black}{\noindent \textbf{Non-ideal factors:} Process, Voltage, Temperature (PVT) variation is captured through threshold voltage distributions with $\sigma_{V_T}$ of 50~mV and 200~mV in the BER and accuracy simulations (Fig.~5). The 200~mV value represents scaled-node FE-NAND variability based on published data~\cite{yoon2023qlc}. Sense margin, IR drop, and leakage in large 3D NAND arrays are inherent to the platform and managed by established array design practices including guard-banding, reference voltage tuning, and bitline precharge schemes. The non-ideal-factor analysis for the identical CUA/CBA in-storage compute architecture has been validated in~\cite{hsu2023storage}, confirming robustness of the in-storage compute approach under these conditions.}

\textcolor{black}{\noindent \textbf{Thermal analysis:} FeFETs operate at significantly lower programming voltages ($<$14~V) compared to conventional CTF NAND ($>$20~V). Since power density scales approximately as $V^2$, the programming power density in FE-NAND is roughly $(14/20)^2 \approx 0.49\times$ that of CTF NAND. Commercial CTF NAND has been validated at $>$200 stacked layers without thermal failures at higher power densities, providing confidence that FE-NAND will have additional thermal margin. Furthermore, in the in-storage compute read-dominant workload, thermal load is governed by sensing operations (sub-$\mu$A currents at low voltages), which contribute negligible power density compared to programming operations.}

\textcolor{black}{\noindent \textbf{Time and energy scaling with logic level:} The energy and latency at each logic level (SLC through QLC) are shown in Supplementary Fig.~S8 for both CTF 3D-NAND and FE-NAND with the same number of WLs per tile. While the per-tile read cost increases with higher logic levels due to the $2^n$ read steps per cell, the number of tiles required to store a fixed-size database decreases by $n\times$ because each tile stores $n\times$ more data. The net total search time therefore scales as $2^n / n$, resulting in only a modest increase from SLC to QLC for the same database.}

\begin{figure}[!h]
    \centering
    \includegraphics[width=0.8\linewidth]{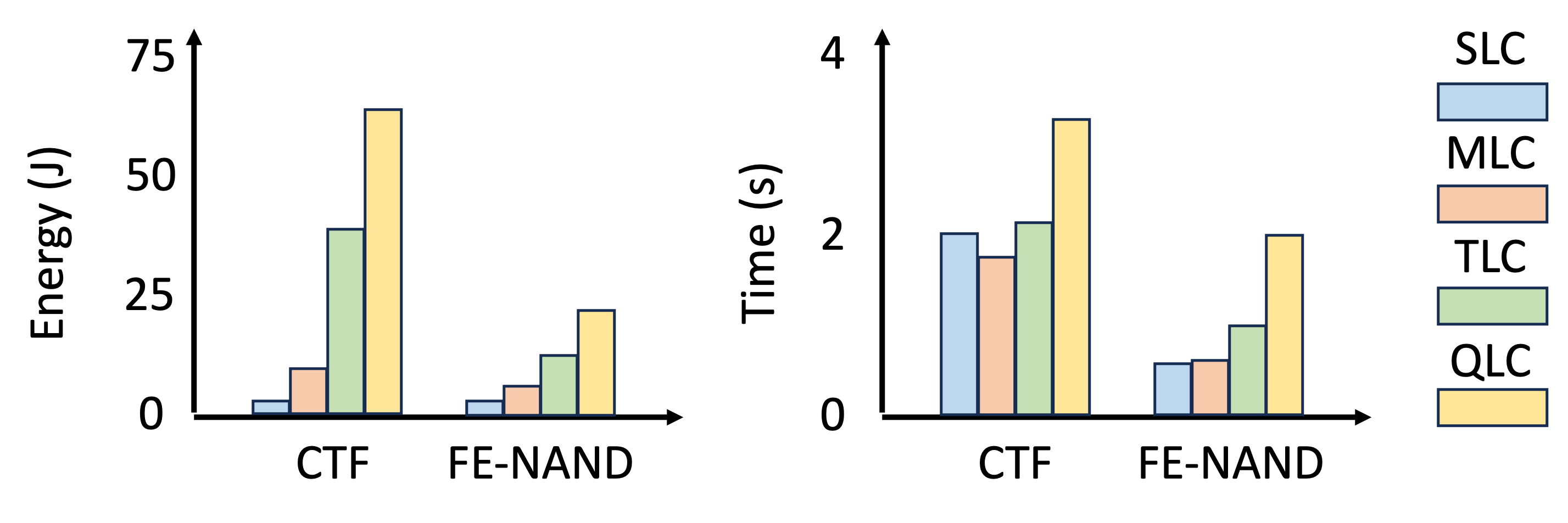}
    \caption{The energy and latency obtained from hardware simulations for conventional CTF 3D-NAND and FE-NAND with equal number of WLs is shown here. } 
    \label{S1}
\end{figure}

\newpage
\section*{Supplementary note 9. Multi-bit operation}

The multi-bit operabilitiy of 3D NAND is determined by the memory window and variation of different $V_T$ states. Multiple factors determine variation in 3D NAND devices including process node (number of WLs), channel materials (poly-Si vs. single-crystal channels), and operating temperature. The threshold margin is estimated by computing the $V_T$ degradation due to data retention, endurance and read noise. Along side hardware considerations, error correction codes also play a critical role. While exact memory window specifications are proprietary, published research and reverse-engineered analyses suggest that conservatively, 7.5 V could enable QLC compatibility in 3D NAND. It is also worth noting that HLC (6-bit) and HeLC (7-bit) operation have been demonstrated at 77 K with 6-8 V MWs by Kioxia \cite{aiba2021cryogenic,aiba2023demonstration,tanaka2022toward,ishimaru2019future}.

\newpage
\textcolor{black}{
\section*{Supplementary note 10. Architectural comparison}}

\textcolor{black}{
\noindent \textbf{Comparison with NOR-type CAM and crossbar architectures.} Supplementary Table~2 summarizes the architectural trade-offs between FE-NAND and alternative compute-in-memory approaches.}

\begin{table*}[!htbp]
\centering
\resizebox{\textwidth}{!}{
\begin{tabular}{lccc}
\hline
\textbf{Attribute} & \textbf{NOR-type FeFET CAM} & \textbf{ReRAM Crossbar} & \textbf{FE-NAND (this work)} \\
\hline
Search complexity & $O(1)$ & $O(1)$ & $O(N)$ \\
Per-cycle parallelism & Full array & Full array & Full page (thousands of BLs) \\
On-chip capacity & $\sim$MB & $\sim$MB (3M cells at 40~nm) & $>$100~TB (3D NAND) \\
Bit-area density & Low & Low--moderate & $>$100~GB/mm$^2$ \\
Readout type & Digital match-line & Analog (ADC) & Digital multi-step \\
Precision issues & Match-line leakage & IR drop, sneak paths, noise & None (digital) \\
Scalability to TB & Not feasible & Not feasible & Inherits 3D NAND \\
Manufacturing maturity & R\&D & R\&D (40~nm) & Production (14~nm equiv.) \\
\hline
\end{tabular}
}
\caption{Architectural comparison of FE-NAND vs.\ NOR-type CAM and crossbar approaches for similarity search workloads.}
\end{table*}
\textcolor{black}{
While NOR-type CAM and crossbar arrays offer $O(1)$ search latency at the array level, their limited on-chip capacity ($\sim$MB) requires repeated off-chip data fetches for TB-scale workloads. The energy and latency cost of data movement dominates the total system budget in these architectures. FE-NAND's $O(N)$ complexity is offset by its ability to store the entire TB-scale database in-storage, eliminating the data-movement bottleneck. As shown in Table~1 of the main text, FE-NAND achieves $>$700$\times$ lower total latency and $\sim$28$\times$ better total energy efficiency than the best-reported ReRAM crossbar solution for the 100~TB proteomics OMS workload.
}

 \newpage
\section*{References}

 \bibliography{references}

%% file: addendum.tex
 \begin{addendum}
 \item[Acknowledgements] 

This work was supported in part by Samsung Electronics (IO-250304-12193-01) and, in part by SUPREME and PRISM, two of the seven SRC-DARPA JUMP 2.0 centers. Fab was done at the IEN, supported by the NSF-NNCI program (ECCS-1542174).

\item[Author Contributions] 

P.V., T.R., S.D. and A.K. conceived the idea. P.V. performed the electrical measurements and analyzed the data. T.H.P. performed array level simulations. P.H., S.P., Z.L. and Z.X. benchmarked the system. C.P., L.F. and H.C. fabricated the devices. P.R., H.J., A.G., H.T. T.S. supported the electrical measurements. W.X., F.P., K.F., W.C., K.K., S.L., K.K., W.K., D.H., D.K., S.Y., M.K. and T.R. participated in useful discussions. All the authors wrote the manuscript.
 
 \item[Competing Interests] The authors declare that they have no competing financial interests.

 \item[Correspondence] 
Correspondence to P. Venkatesan and S. Dutta.
 \end{addendum}

%% file: methods.tex
\renewcommand{\figurename}{Extended Data Fig.}
 \setcounter{figure}{0}
\renewcommand{\tablename}{Extended Data Table.}
 \setcounter{table}{0}
\clearpage
\begin{methods}
\subsection{Device fabrication}

The detailed process flow for FEFETs is described in Supplementary note 1. The Hf\textsubscript{0.5}Zr\textsubscript{0.5}O\textsubscript{2}
and Al\textsubscript{2}O\textsubscript{3} layers in the ferroelectric gate stack are deposited with thermal ALD at 250$^\circ$C. Both the standard 19 nm FEFET and engineered FEFET (8 nm HZO/ 3 nm Al\textsubscript{2}O\textsubscript{3}/ 8 nm HZO) are were annealed at 500$^\circ$C for 30 seconds. Al\textsubscript{2}O\textsubscript{3} is chosen as the tunnel dielectric layer based on extensive design space exploration presented in \cite{fernandes2024material,fernandes2024optimizing}. Tunnel dielectric layer, where the dielectric is inserted in the middle of the ferroelectric gate stack, is chosen over a gate blocking layer, where a dielectric is inserted next to the gate, due to their superior retention performance \cite{venkatesan2024demonstration,venkatesan2025enhanced}. All our experiments have been performed on devices of width 10 $\mu m$ and length 3 $\mu m$. 

\subsection{Electrical characterization}  

The basic electrical characterization like pulsed transfer characteristics and drain currents for SLC and MLC dot products were measured using Keysight Semiconductor Parameter Analyzer B1500 equipped with B1530 WGFMUs and B1511 Medium Power SMUs. A Keysight Low-leakage Switch Matrix B2200 was used for characterizing the FE-NAND strings. \textcolor{black}{A current compliance of 100~$\mu$A was applied during all DC spot read measurements. All single-cell dot product measurements (Fig.~3) were performed on fabricated NAND strings with unselected wordlines biased at $V_{pass} = 5$~V, so that pass-cell series resistance effects are inherently included in the experimental data.}

\subsection{Array level simulations}  

The virtual source model is employed to simulate the FET, while the Preisach model is utilized to capture the ferroelectric switching kinetics of HZO. The governing equations and parameters for both models are provided in Fig. S6 along with the calibration of the models with experimental $I_D - V_G$ data. 

\subsection{System benchmarking}  

To evaluate the system-level performance of Fe-NAND and CTF NAND, we utilize the proposed in-storage processing architecture that incorporates both CMOS-under-array (CUA) and CMOS-bonded-with-array (CBA) 3D heterogeneous integration techniques, achieving compact design \cite{hsu2023storage}. High-voltage circuits including word-line switching matrices and pass transistors are fabricated beneath the 3D array using the CUA method at a 65 nm logic-equivalent node, ensuring reliable high-voltage programming and erasure capabilities. On the other hand, low-voltage components including near-memory HD encoders, digital logic, buffers, decoders, and analog-to-digital converters (ADCs) are deployed on a separate CMOS wafer at a 7 nm node. These low-voltage circuits were then bonded face-to-face with the 3D memory wafer using Cu-Cu hybrid bonding, with an inter-tier pitch of approximately 1 $\mu$m \cite{chen20203d}, facilitating high-bandwidth data transfer. The implemented near-memory HD encoder and search circuits are synthesized using the ASAP 7 nm process-design-kit (PDK) \cite{clark2016asap7}. Peripheral circuits for the 3D arrays were adapted from NeuroSim \cite{peng2019dnn+}, and the system operated at a clock frequency of 1 GHz.

\textcolor{black}{System performance is modeled hierarchically: experimentally calibrated Preisach + Virtual Source SPICE models (Supplementary Note~6) provide unit-cell FeFET parameters; array-level RC parasitics (bitline resistance, wordline RC delay) are based on industry-grade 3D NAND chip-level parameters at a 14~nm equivalent node; peripheral circuits (row decoders, sense amplifiers, page buffer latches, ADCs) are extracted from NeuroSim v1.5, which has been validated against silicon data for NAND-based architectures; and system-level energy and latency are computed using cycle-accurate simulation that accounts for all data-movement, compute, and peripheral overhead within the CUA/CBA IMC architecture. Non-ideal factors including PVT variation ($\sigma_{V_T}$ = 50--200~mV), sense margin, IR drop, and leakage are incorporated through the $V_T$ distributions used in BER and accuracy calculations (Fig.~5), and by adopting the validated non-ideal-factor analysis for the identical CUA/CBA IMC architecture from~\cite{hsu2023storage}.}

We compare our approach against HyperOMS on GPU\cite{kang2022massively} and  RapidOMS\cite{pinge2024rapidoms} on FPGA and Process-In-Memory (PiM) accelerators including DRAM\cite{kang2024dram}, ReRAM\cite{fan2024efficient} and 3D NAND\cite{hsu2023storage}. For HyperOMS and RapidOMS, energy and latency are measured through experiments on RTX4090 GPU and Samsung SmartSSD FPGA. For PIM accelerators, the benchmark results are obtained through in-house cycle-accurate simulators based on device characteristics reported in the literature \cite{fan2024efficient,fan2024specpcm,hsu2023storage}. RRAM benchmarks were computed at 130 nm node and scaled down to the 40 nm node using the method proposed by Stillmaker et al \cite{stillmaker2017scaling}.

\clearpage

\end{methods}